\documentclass[10pt]{article}
\pdfoutput=1

\usepackage[paper=a4paper,dvips,top=1.5cm,left=1.5cm,right=1.5cm,
    foot=1cm,bottom=1.5cm]{geometry}
    
\usepackage{graphicx}
\usepackage{graphics}
\usepackage{amsmath}
\usepackage[latin1]{inputenc}
\usepackage{braket}
\usepackage{textcomp}
\usepackage{amsmath}
\usepackage{amssymb}
\usepackage{multicol}
\usepackage{multirow}
\usepackage{caption}
\usepackage{lscape}
\usepackage{color}
\usepackage{wrapfig}
\usepackage{fancyhdr}
\usepackage{mdframed}
\usepackage{bm}
\usepackage{longtable}
\usepackage[table]{xcolor}
\usepackage{bbm}

\usepackage{ctable}
\usepackage{makecell}
\usepackage{array}
\newcolumntype{L}[1]{>{\raggedright\let\newline\\\arraybackslash\hspace{0pt}}m{#1}}
\newcolumntype{C}[1]{>{\centering\let\newline\\\arraybackslash\hspace{0pt}}m{#1}}
\newcolumntype{R}[1]{>{\raggedleft\let\newline\\\arraybackslash\hspace{0pt}}m{#1}}

\usepackage{hyperref}

\title{
\rule{16cm}{2.5pt}\\
\vspace*{.1cm}
\textbf{\Large{Netconomics:}\\[.3cm]
		\large{Novel Forecasting Techniques from the Combination of\\Big Data, Network Science and Economics}}\\[-.1cm]
\rule{16cm}{2.5pt}
}
\author{
        \textsc{$\mbox{Andreas \textsc{Joseph}}^{1,2\,\thanks{\href{mailto:andreasj@bu.edu}{Correspondence to: \texttt{andreasj@bu.edu}}}}$},
        \textsc{$\mbox{Irena \textsc{Vodenska}}^{2,3}$},
           \textsc{$\mbox{Eugene \textsc{Stanley}}^{2}$},
        \textsc{$\mbox{Guanrong \textsc{Chen}}^1$}\\
        \mbox{}\\
         \small{$\mbox{ }^1$ Centre for Chaos and Complex Networks, Department of Electronic Engineering,} \\ 
         \small{City University of Hong Kong, Hong Kong S.A.R., China} \\
         \small{$\mbox{ }^2$ Center for Polymer Studies and Department of Physics, Boston University, Boston, MA 02215, USA}\\
         \small{$\mbox{ }^3$ Administrative Sciences Department, Metropolitan College, Boston University, Boston, MA 02215 USA}\\
}

\begin{document}
%
\thispagestyle{empty}
\maketitle
\textbf{The combination of the network theoretic approach with recently available abundant economic data leads to the development of novel analytic and computational tools for modelling and forecasting key economic indicators. The main idea is to introduce a topological component into the analysis, taking into account consistently all higher-order interactions. We present three basic methodologies to demonstrate different approaches to harness the resulting network gain. First, a multiple linear regression optimisation algorithm is used to generate a relational network between individual components of national balance of payment accounts. This model describes annual statistics with a high accuracy and delivers good forecasts for the majority of indicators. Second, an early-warning mechanism for global financial crises is presented, which combines network measures with standard economic indicators. From the analysis of the cross-border portfolio investment network of long-term debt securities, the proliferation of a wide range of over-the-counter-traded financial derivative products, such as credit default swaps, can be described in terms of gross-market values and notional outstanding amounts,
which are associated with increased levels of market interdependence and systemic risk. Third, considering the flow-network of goods traded between G-20 economies, network statistics provide better proxies for key economic measures than conventional indicators. For example, it is shown that a country's gate-keeping potential, as a measure for local power, projects its annual change of GDP generally far better than the volume of its imports or exports.}\\
\tableofcontents
\clearpage
\section{Introduction}
\label{sec:intro}
Since the latest global financial crisis 2007-2009 (GFC'08) and the resulting European Sovereign Debt Crisis (ESDC), which has not yet been resolved, policy makers, academia and the public have become aware of the strong and important interconnectedness and interdependence of the global financial and economic architecture \cite{net_opp,crotty,preis,cds_risk}. Additionally, standard techniques in macroeconomics partly failed to describe or foresee these major downturns, as pointed out by Jean-Claude Trichet in his opening address at the ECB Central Banking Conference (Frankfurt, 18 November 2010): ``Macro models failed to predict the crisis and seemed incapable of explaining what was happening to the economy in a convincing manner. As a policy-maker during the crisis, I found the available models of limited help. In fact, I would go further: in the face of the crisis, we felt abandoned by conventional tools.''
\begin{wrapfigure}{r}{0.6\textwidth}
  \vspace{-10pt}
  \begin{mdframed}
  \begin{center}
    \includegraphics[width=1.\textwidth]{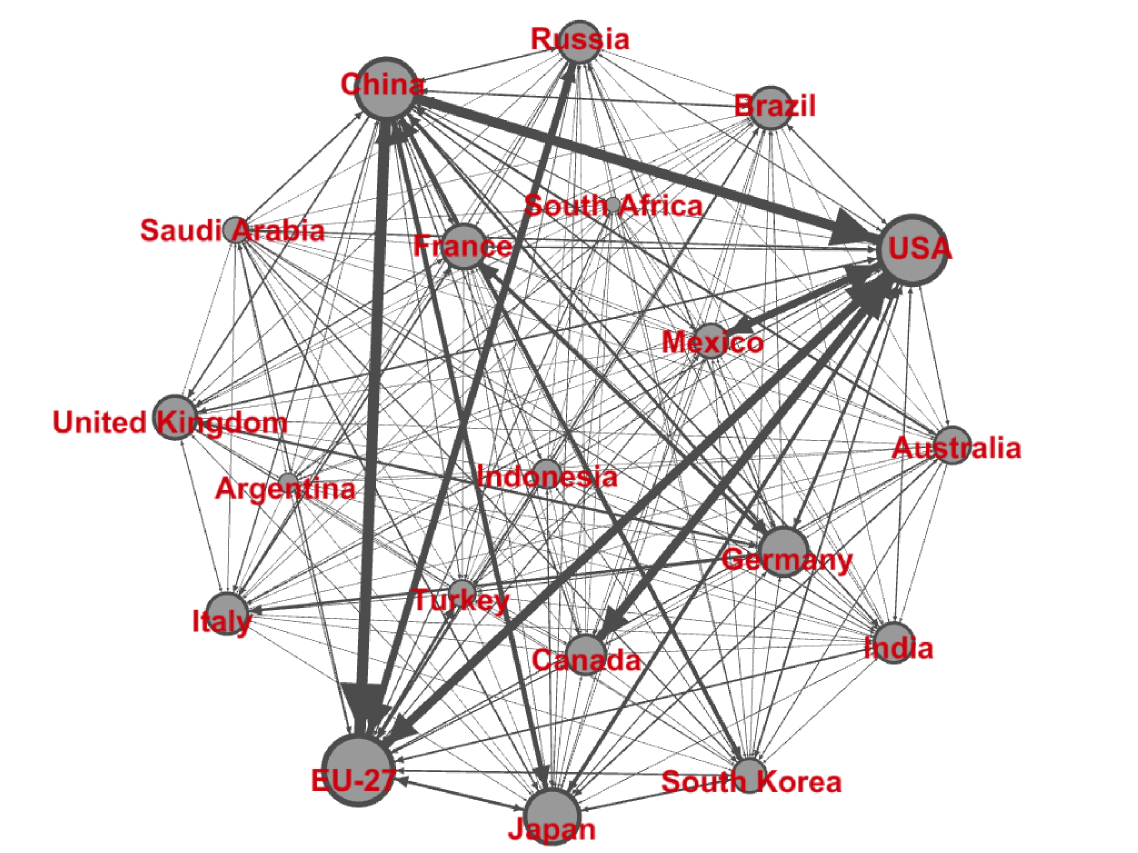}
  \end{center}
  \vspace{-5pt}
  \caption{\small{The \textbf{International trade network of G-20 countries} (ITN-20). Directed edges (links) between any two nodes (countries) are given by the total amount of trade between these countries. The thickness of arrows and the sizes of their heads indicate the magnitudes of trade flows in both directions. Node sizes are set according the logarithm of a country's GDP. One can clearly see how the nodes Europe, China and the United States form a major global triangle of trade and how other countries connect to it.}}
  \end{mdframed}
  \vspace{-10pt}
\label{fig:g20_wtw}
\end{wrapfigure}
\noindent
However, recent years have also seen the rise of \textit{Big Data}, i.e.\ the availability of large amounts of high-quality digitalised data, as well as the development of \textit{network science}, which investigates the properties of systems composed of a large number of connected components. The combination of Big Data and network science offers a vast number of potential applications for the design of data-driven analysis and regulation tools, covering many parts of the society and academia \cite{futurict}. Especially, \textit{econometrics} is a field where merging current analysis techniques with data- and network science is expected to provide large gains to modern economics.\\
The main idea behind network science is to generate a \textit{network gain} by consistently considering multiple, as well as higher-order interactions between the individual components of a large system, such as users in a social network, banks in the financial market, or the interaction of whole economies on a global scale. There are different directions from which network science can be coupled to Big Data and be applied to econometrics, where the main differences consist in the way \textit{network} is generated. The two major approaches, which will be presented in this report, are the consideration of a physical structure, as for example the flow of goods or the transaction of money between certain agents in a market, and a relational connection between two elements of a larger group, where one might think of the correlated changes of stock prices in a portfolio.\\ 
An informative example for the first approach to constructing a network from data and to achieving a qualitative network gain is the international trade network of G-20 countries (ITN-20), as visualised in \hyperref[fig:g20_wtw]{Fig.\ 1}. Every node represents a country (or a group of countries in the case of the EU-27 countries), where sizes are set according to the logarithm of a country's GDP (gross-domestic product). Edges (links) between any two countries represent the trade flow between them, while the thickness and the size of arrow heads indicate the magnitudes of the total trade volume and the directional components, respectively. One can clearly identify a global triangle of trade, between Europe, China and the United States, and how the remaining countries connect to it. This rather simple example illustrates already how the network approach delivers a bird's eye view of the intrinsic complexity, which can be readily read out from the figure.\\
In this report, we present three case studies from macroeconomics where the network picture had been merged with conventional analysis tools and market indicators, facilitating the fusion of tested methodologies with refreshingly new approaches.\\
In the \hyperref[sec:mlr]{first case}, we construct an interrelational structure (network) between trading and financial indicators (nodes) by the means of a multiple linear regression (MLR) algorithm, which links every indicator to one or several other indicators (edges). The resulting relational network will be called \textit{Global Balance of Payments Network} (GBoPN). The network itself represents an evolution framework which describes, and even forecasts, most of the indicators with high accuracy.\\
In the \hyperref[sec:pin]{second case}, global \textit{Portfolio Investment Networks} (PIN) are constructed from cross-border position of equity and debt securities \cite{pin}. We will demonstrate how network statistics can be used to model the proliferation of a wide range of over-the-counter-traded financial derivative products, which have been identified as playing a major role during the GFC'08. Combining the derived phenomenological model with macroeconomic market data allows to design an early-warning mechanism for future financial crises.\\
In the \hyperref[sec:g20]{last example}, we come back to the ITN-20 and demonstrate how local node centralities\footnote{``centrality'' is the general term for a large variety of node and edge measures in graph theory.}, i.e.\ network measures which quantify a node's relations to its nearest neighbours, can be used to project standard economic indicators, such as the annual change of GDP, i.e.\ economic growth.
\section{The MLR-Fit Network of Global Balance of Payment Accounts}
\label{sec:mlr}
\begin{wrapfigure}{r}{0.3\textwidth}
  \vspace{10pt}
  \begin{mdframed}
    \includegraphics[width=1\textwidth]{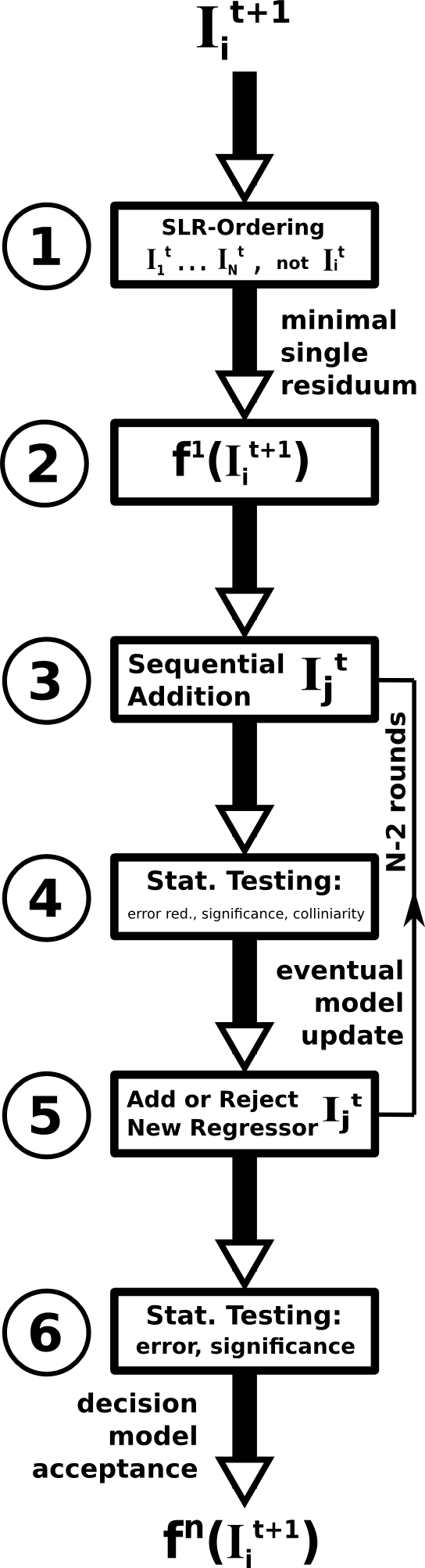}
  \vspace{-15pt}
  \caption{\small{Flow diagram of \textbf{iterative MLR-fit algorithm} for each indicator $I_i(t+1)$ to determine the coefficient matrix $\bm{\beta}$. See text for details.}}
  \end{mdframed}
  \vspace{-10pt}
\label{fig:algo}
\end{wrapfigure}

Especially, since the GFC'08 and the ESDC, the understanding of economic contagion has attracted growing interest, aiming to explain and measure the spreading of economic downturns between countries and across asset classes \cite{contagion1,contagion2,preis,jpm_cross}. The modelling of global economic interactions between the different macro-components originating from multiple countries faces substantial difficulties due to the large number of possible interaction channels, i.e.\ due to the underlying computational complexity of the problem.
\begin{wrapfigure}{l}{0.4\textwidth}
  \vspace{-10pt}
  \begin{mdframed}
  \begin{center}
    \includegraphics[width=1.\textwidth]{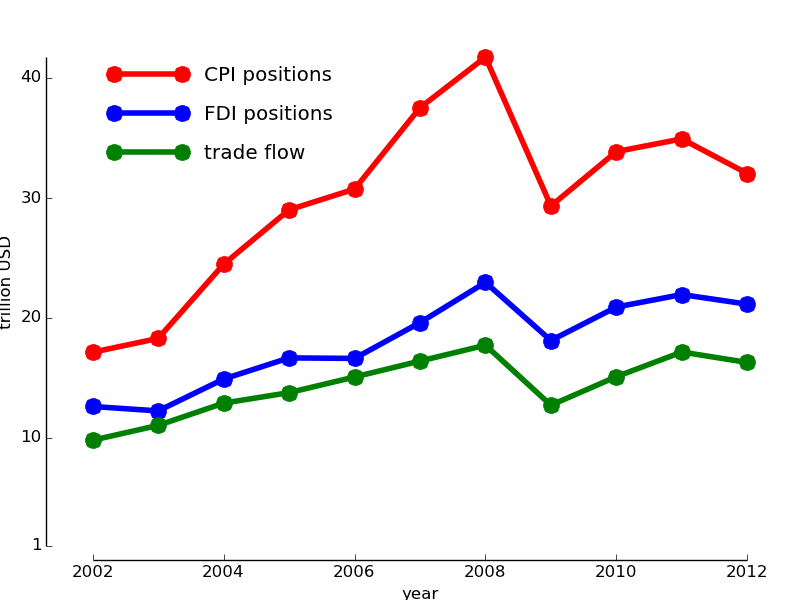}
  \end{center}
  \vspace{-5pt}
  \caption{\small{Comparison of the \textbf{magnitudes of global trade flows and investment positions} (CPI and FDI) between 2002-2012. Because the total trade flow within one year tracks well the start-of-the-year positions of both types of investment, we consider a combination of trade flows and investment positions.}}
  \end{mdframed}
  \vspace{-10pt}
\label{fig:vols}
\end{wrapfigure}
\noindent
In this section, we present how a standard technique from economic analysis, namely MLR, can be combined with Big Data and network science to tackle the above-stated problem by delivering an accurate phenomenological description with a high predictive power. We note from the beginning that the applied methodology is not confined to any particular field, but has a large number of potential applications, whenever the criterion of sufficient data availability is fulfilled.\\
For a total of 60 countries during eleven consecutive years (2002-2012), we construct a relational network between eight financial and trade indicators of each country. These constitute major parts of a country's balance of payments. The eight indicators are the trade of goods \cite{comtrade}, foreign direct investment (FDI) \cite{unctad} and cross-border portfolio investment (CPI) of equity- and debt securities \cite{cpis} on the nationally aggregated level, where we differentiate between in- and outbound relations. In \hyperref[fig:vols]{Fig.\ 2}, it is shown that the aggregated global trade flow within one year tracks the corresponding start-of-the-year investment positions, CPI and FDI. This is the reason why we will focus on a country's total exports and imports, as well as in- and outbound investment positions, in this analysis. Example indicators, which are included in this analysis, are the total exports of China (China: Exports) and the aggregated foreign holdings of US debt securities (USA: Debt (in)). All indicators have been adjusted for yearly changes in GDP, using the global GDP-deflator \cite{wb_data} (constant year-2012 values).\\
In algebraic terms, a network can be represented by its \textit{adjacency matrix} $\mathcal{A}$, where the element $a_{ij}\neq 0$ if there is an edge between node $i$ and node $j$. One distinguishes between different classes of networks, depending on the possible values the matrix elements $a_{ij}$ are allowed to take. If $a_{ij}\in\left\lbrace 0,1\right\rbrace$, $a_{ij}=a_{ji}$ and $a_{ii}=0$, the resulting network is called a simple graph, which captures the topological structure of the network in which connections between nodes are symmetric. If it only requires $a_{ij}\in\left\lbrace 0,1\right\rbrace$, one obtains a directed graph, where connections between nodes are not required to be reciprocal. In the most general network,  $a_{ij}\in\mathbb{R}$, meaning that one can have asymmetric connections between any two nodes accounting for possibly negative feedback. This is the kind of networks we will consider throughout this report.\\
We are now ready to construct the MLR-fit network between single elements of global balance of payment accounts, which represent the nodes of the GBoPN. The underlying idea of constructing this network is that global financial and trade statistics are related to each other, such that it is possible to find a set of indicators (regressors) which best describes \textit{another indicator} (regressand), using MLR.\\
Let $\vec{I}(t)$ denote the set of all indicators of all countries at time $t$, e.g. in 2007. We are interested in finding the best-fit network coefficient matrix $\bm{\beta}=\beta_{ij}$ which describes $\vec{I}(t+1)$. This can be written in terms of a linear matrix equation
\begin{equation}
\label{eq:mlr}
I_i\left(t+1\right)\,\stackrel{\text{\tiny MLR}}{=}\,\sum_{j=1,j\neq i}^{N}\,\beta_{ij}\,I_j\left(t\right)\,+\,c_i\,,
\end{equation}
where $N$ is the total number of indicators from all countries and $c_i$ is the intercept of indicator $I_i$. A link between indicators $I_i$ and $I_j$ is established if $\beta_{ij}\neq 0$. Note that the model (\ref{eq:mlr}) is highly appealing from a mathematical point of view because of its simplicity, as well as its built-in predictive power, since the coefficient matrix $\bm{\beta}$ can be interpreted as an \textit{evolution operator}, which takes the indicator vector $\vec{I}$ from $t\rightarrow t+1$.\\
However, finding the optimal $\bm{\beta}$ according to the set statistical criteria on the MLR-fit is a computationally hard problem because the number of possible solutions is growing super-exponentially with the number of indicators. That is the point where Big Data and computational algorithms enter the game. Finding a \text{good} solution to this problem is much more likely with the availability of a large amount of data, while an efficient search method will considerably shorten the time to find a solution. We use an interative least-square algorithm, which is based on the assumption that a regressor which individually describes a regressand well (simple regression) is likely to be contained in a group of regressors (multiple regression). The working of the optimisation algorithm is depicted in \hyperref[fig:algo]{Fig.\ 3}: On each $I_i\left(t+1\right)$ time series, perform a simple linear regression (SLR) with each \textit{other} time-lagged time series $I_j\left(t\right)$ (Step\ 1). 

\noindent

\begin{figure}[t!]
  \begin{mdframed}
  \begin{center}
    \includegraphics[width=1.\textwidth]{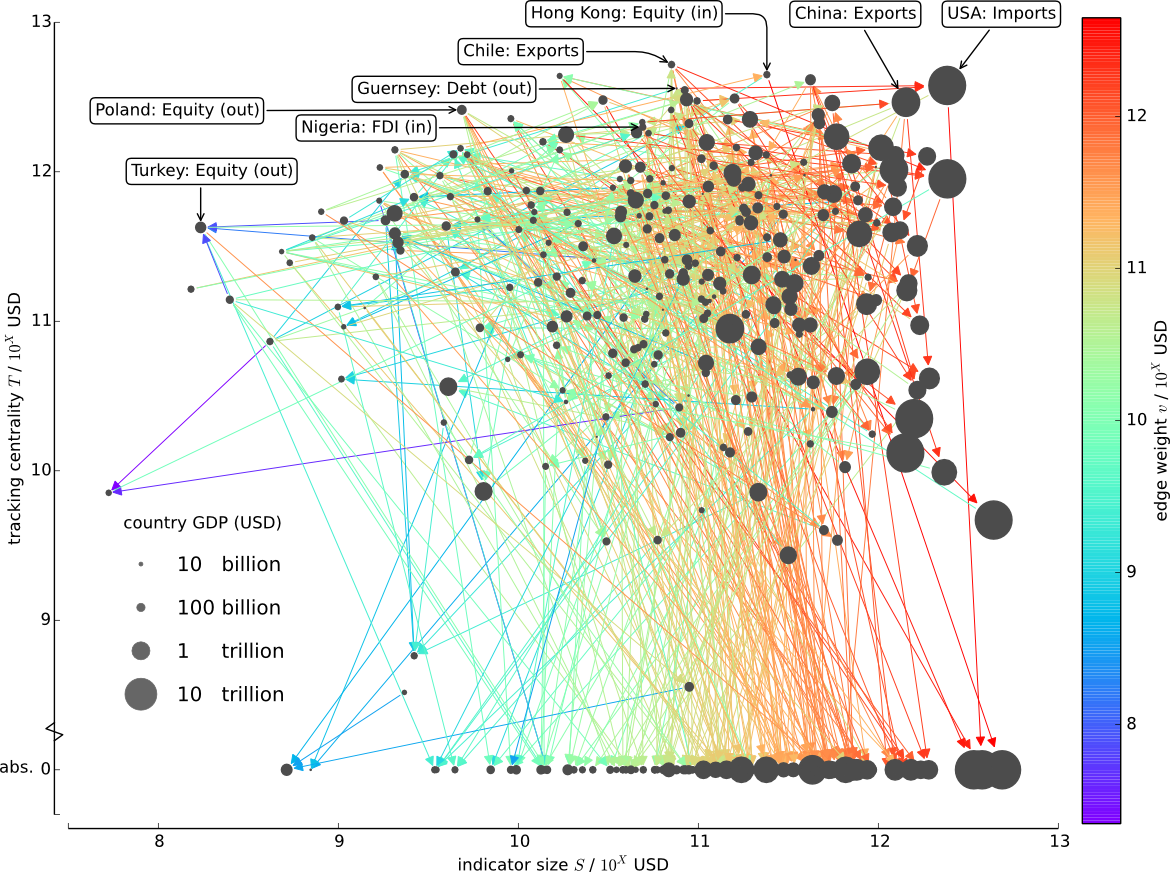}
  \end{center}
  \caption{\small{Visualisation of the \textbf{Global Balance of Payments Network} (GBoPN), constructed with the MLR-fit algorithm shown in \hyperref[fig:algo]{Fig.\ 3}. Indicators (nodes) are represented by dots with sizes corresponding to countries' GDP. The x- and y-positions of each node are set according to the  indicator's time-averaged size $S$ and tracking centrality $T$, respectively. Directed edges, pointing from a regressor $i$ to its regressand(s) $j$, are colour-coded according  to the tracked value $v_{ij}$. A key-result is the high tracking capabilities of many indicators of small sizes or from small economies, which might be used as proxies or ``thermometers'' for larger economic changes. Highlighted examples include the outbound equity securities positions of Poland or the outbound debt securities positions of Guernsey. Indicators which have not been identified as statistically significant regressors, but are tracked by others, are positioned on the x-axis.}}
  \end{mdframed}
\label{fig:bop}
\end{figure}

Pick the regressor $I_j$ which generates the smallest error (residuum) for the starting model $f^1\left(I_i^{t+1}\right)$ (Step\ 2). Add additional regressors to the model, according to the ordering of residua from Step\ 1 (smallest to largest), and test the resulting model for error reduction, statistical significance of all regressors (t-test) and collinearity (variance inflation factor (VIF) and condition number of the normalised design matrix, Steps\ 3 and\ 4). Based on the test results, update the model (or not) and go back to Step\ 3. Repeat this step for $N-2$\ times (Step\ 5). Test the statistical significance (F-test) and error of the final model $f^n\left(I_i^{t+1}\right)$, where $n$ is the final number of regressors, then accept or reject it (Step\ 6).\\
It yields one row of the coefficient matrix $\bm{\beta}$ for each indicator $I_i\left(t+1\right)$. The final result will crucially depend on the chosen requirements on the maximally allowable error, the statistical significance of each coefficient and on collinearity bounds, where there is a general trade off between the maximal error on the one side and the statistical significance and collinearity between regressors on the other side. Depending on the achievable balance between these quantities, the MLR-fit model as a whole may be accepted or rejected.\\
A few words need to be said about the selection of countries and the availability of data. The 60 countries were chosen according to a $95\%$-criterion on the cumulative amount of the total monetary value (in USD) of all eight indicators taken together. Meanwhile the union of countries is chosen, which hold at least $95\%$ of the average total value of each indicator. Theoretically, this makes a total of $60\times 8=480$ indicators. Unfortunately, data are partly incomplete, so that we are left with a total of $405$ indicators, or about $84\%$ of the expected number. This is still a considerable number and enough data to achieve good fits for the great majority of indicators, as we will see in a moment. Rather strict statistical criteria had been set in order to accept or reject each single fit at each step, namely, a significance level for the t- and F- tests of $\alpha=2.5\%$, a maximal time-averaged final fit error of each $I_i$ of $10\%$ and a maximal condition number and VIF on the design matrix of ten and five, respectively.\\
In order to test the forecasting capability of the MLR-network model (\ref{eq:mlr}), we remove the year-2012 data before performing the fit to use it for an out-of-sample test later. Taking the one-year time shift between regressor and regressand variables into account, we are left with a series of nine data points to do the fitting, which will turn out to be enough to do proper now- and forecasting. \\
The final fit results in an acceptable model for $372$ of the $405$ indicators with an average error of $2.9\%$ and a median error of $1.7\%$. The detailed MLR-fit results for each indicator are summarised in \hyperref[tab:mlr_stats]{Tab.\ 1} in the \hyperref[sec:app]{Appendix}. The resulting GBoPN generated from the coefficient matrix $\bm{\beta}$, where an edge from indicator $i$ to indicator $j$ is drawn if $\beta_{ij}\neq 0$, which turns out to be connected, meaning that it cannot be separated without cutting some edges. This is the first non-trivial insight from the network perspective, since it tells us that any of the initial $405$ indicators is significantly coupled to at least one other indicator and that the resulting network covers all indicators for every of the 60 countries. Especially, it doesn't show a separation into geographically localised clusters, which could have been a valid expected outcome, but the picture of a globally interacting multi-layered economic system is drawn. A feature which strongly underlines the importance of cross-border economic relations is that basically all best-fit edges (more than $98.4\%$) are connecting indicators from different countries, while about one quarter of them connects indicators from different classes, such as trade and FDI. \\
The GBoPN turns out to be extremely sparse with an edge density\footnote{the number of maximally possible edges $N(N-1)$ dived by the number of actual edges, which, in our case, is the number of non-zero coefficient $\beta_{ij}$.} of less than $0.5\%$. This means that a very small number of relations between all possible pairs of indicators is enough to describe nearly all indicators with high accuracy. A great majority of $298$ indicators are \textit{tracked} by two indicators or even just one. On the other side of the spectrum, there is a smaller number of indicators which \textit{track} a large number of other indicators. This, in the language of graph theory means that they have a large out-degree. These indicators might prove potentially useful for the purpose of macroeconomic monitoring and forecasting because their observation is expected to deliver over-proportional information about other indicators and their host-countries. An additional criterion for identifying nodes of interest is the monetary value of indicators they point to, because these vary by several orders of magnitude. To account for the number and size (in terms of their time-averaged values in USD) of a node's regressands, we define its \textit{tracking centrality} as
\begin{equation}
\label{eq:track}
T_i\,\equiv\, \sum_{j=1,j\neq i}^{N}\,\sqrt{R^2_{ij}}\,S_j\,,
\end{equation}
were $R^2_{ij}$ is the \textit{coefficient of determination} between indicators $i$ and $j$, and $S_j$ is the time-averaged monetary value of indicator $i$. By definition, $R^2_{ij}\neq 0$, whenever $\beta_{ij}\neq 0$. $\sqrt{R^2_{ij}}$ equals the absolute value of the Pearson product correlation coefficient between indicators $i$ and $j$ and measures the fraction of variation which is mutually described.\\
A comprehensive visualisation of the entire GBoPN is shown in \hyperref[fig:bop]{Fig.\ 4}. Each indicator is represented by a dot of the size of its host-country's GDP. The x-position of each node is its size $S$ and the y-position is its tracking centrality $T$, where a higher value means that a larger amount of economic value is quantitatively described by this indicator. The edges of the GBoPN are given through colour-coded arrows pointing from a regressor to its regressand(s). The colour is set according to the relation $v_{ij}=\sqrt{R^2_{ij}}\,S_j$, quantifying the value of this relationship. As such, the tracking centrality of an indicator is given by the values of its outbound connections. A key-result from \hyperref[fig:bop]{Fig.\ 4} is the high tracking capabilities of many indicators of small sizes or from small economies, which might, in turn, be used as proxies or ``thermometers'' for larger economic changes. Such nodes can be easily spotted as having strong connection spanning a large horizontal distance (several orders of magnitude) from left to right. Highlighted examples include the outbound equity securities positions of Poland or the outbound debt securities positions of Guernsey. We note that the so-called off-shore financial centres \cite{def_ofc} have high tracking centralities on average, due to their inherently strong coupling to the financial system.\\
Given the small average fit error, it is possible to make explicit use of the network structure, taking higher-order interactions into account, and track indicators over short paths, where the averaged errors are small. The maximally expected error is then the sum over all contributing errors along this path. A simple but instructive example is the tracking of Spain's inbound FDI, using Greece's imports with an aggregated error of $4.5\%$ over 2002-2011, which is the only regressor in this case.\\
Another very interesting application of the MLR-fit model (\ref{eq:mlr}) is the forecasting of indicators. As mentioned, we excluded the year-2012 data when performing the fit in order to perform an out-of-sample test of the time evolution operator $\bm{\beta}$ (in combination with the intercepts $\vec{c}$). The results of this test are summarised in the last columns of \hyperref[tab:mlr_stats]{Tab.\ 1}. Besides having some outliers, the median forecasting error for the $372$ macroeconomic indicators turns out to be $8.5\%$ with an standard deviation of $15.5\%$, excluding those 3 indicators with an forecasting error of more than $100\%$. These now- and forecasting capabilities are especially powerful for international trade, because trade flows are observed to lag somehow behind cross-border investment positions, as seen in \hyperref[fig:vols]{Fig\ 2}. Among many others, this is the case for the imports and exports of the United States, Germany and China, which greatly enhances the descriptive power for these major indicators. Even for the case that such a relation cannot be found, there is a considerable likelihood that a ``good second-best'' fits, involving such a network configuration. The same is true for nodes lying on the x-axis of \hyperref[fig:bop]{Fig.\ 4}, which are expected to have certain tracking capabilities, but, in most cases, have been eliminated by the statistical requirements on the final model fit.\\
The potentially large descriptive and predictive capabilities of the MLR-network model (\ref{eq:mlr}) are not meant to stand alone, but are expected to provide additional insights in combination with country- and indicator-specific information, which we have neglected with intention to focus on the applicability of data- and network mining approaches.\\
\vspace*{10cm}
\clearpage
\section{Portfolio Investment Networks and Indicators for Financial Crises}
\label{sec:pin}
We consider large-scale financial networks, which are derived from an actual physical layer, namely, nationally aggregated cross-border holdings of portfolio investment  in equity and debt securities \cite{cpis,pin}. The resulting networks are called portfolio investment networks (PIN). As has been demonstrated in \cite{pin}, their properties can be used as proxies to measure the structural robustness of the global financial system, as seen from the network perspective, and the interdependence of financial markets, taken by the proliferation of financial derivative products.\\
In this section, we demonstrate how the combination of network statistics and market data can be used for the design of an early-warning mechanism for global financial crises. We measure the level of interconnectedness of long-term debt securities holdings and relate them to the aggregated market values, in terms of gross-market values (GMV) and notional outstanding amounts (NOA), of a wide range of financial derivative products, such as credit default swaps (CDS) and equity- and commodity-linked derivatives (E- and CLD). The underlying rationale is that debt securities can by definition be traded, while potentially being bundled-up for the formation of higher-order derivative products. This is expected to lead to an increase of cross-border investment holdings, if these products are sold on an international market or if there are foreign market participants, which is generally the case.\\
\begin{figure}[h!]
  \begin{mdframed}
  \begin{center}
    \includegraphics[width=1.\textwidth]{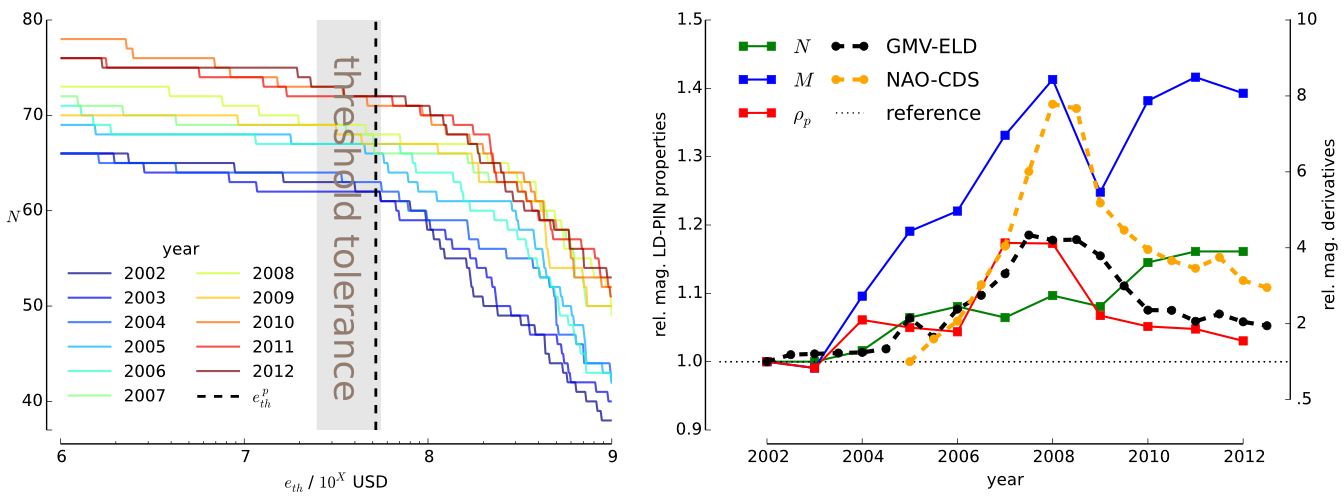}
  \end{center}
  \caption{\small{Key properties of the \textbf{Long-term Debt Portfolio Investment Network} (LD-PIN). Left: Edge threshold dependence of the largest strongly-connected component. Removing stronger and stronger links, one finds a universal percolation point at about 52\ million\ USD, beyond which the whole network disintegrates quickly. The level of connectivity at this point is expected to contribute dominantly to the global properties of the LD-PIN. The shaded area indicates the threshold tolerance of the phenomenological model (\ref{eq:mm}) for the description of market values of financial derivative products. Right: Temporal evolution of the number of nodes $N$ (countries), number of edges $M$ (cross-border positions), the resulting edge density at the percolation point $\rho_p$, gross-market value of equity-linked derivatives (GMV-ELD) and notional outstanding amounts of credit default swaps (NOA-CDS). All quantities are given with respect to their year-2002 values. Source:\ \cite{pin}.}}
  \end{mdframed}
\label{fig:ld_props}
\end{figure}

\noindent
For the case of debt securities, the insurance of higher-order products is expected to be coupled to the proliferation of CDS. While cross-asset class correlations, originating from certain market mechanisms \cite{jpm_cross}, are expected to translate such increases to other asset- and derivative classes. This is exactly the kind of scenario, which had been identified as leading to the recent GFC'08 \cite{cds_risk,crotty,allison}. Whereas mainly originating from the United States, which plays a leading role in international financial markets, large amounts of securitarised debt obligations had been sold, which created a global network of strong interdependence. Considering the effect of financial derivatives (of any kind) from a dynamical network point of view, high levels of proliferation would naturally be associated with higher \textit{systemic risk}, since there are more channels for contagion and an increased likelihood of catastrophic cascades. For CDS, a larger number of creditors get indirectly coupled to a larger number of debtors via large insurers. During ``good'' times, this might provide better risk protection through diversification. However, an aggravation through collective behaviour may be the outcome during downturns \cite{preis,manias}.\\
Our network set-up is the following. We consider the long-term debt securities (LD-) PIN \cite{cpis} from 2002-2012, covering the GFC'08. A node of the network is given by one of 78 CPIS-reporting countries and a connection from country $i$ to country $j$ (edge) by the aggregated holdings of country $i$ (by its residents and institutions) of LD-securities originated in country $j$. To focus on the dominant global properties of the LD-PIN, we consider the network at its percolation edge threshold of $e_{th}=52$\ million USD, meaning that we remove all edges which are weaker than this value, while we account for changes in world-GDP by using the GDP deflator \cite{wb_data} (constant year-2013 values). We stay with the largest remaining (strongly-)connected component after applying the threshold. Above $e_{th}$, the LD-PIN disintegrates rapidly, meaning that global properties are assumed to be best described at this level of connectivity, because adding weaker edges does not lead to a substantial growth of the network. Note that it is not guaranteed that such a percolation point exists. A continuous disintegration of a network with rising edge threshold can also be the case. The existence of such a ``working point'' is a rather peculiar property of the LD-PIN. The percolation properties of the LD-PIN are depicted on the left of \hyperref[fig:ld_props]{Fig.\ 5}, where the sizes of the largest connected components for values of $e_{th}$ between 1\ million and 1\ billion USD are shown with respect to the number of nodes $N$.\\
The right side of \hyperref[fig:ld_props]{Fig.\ 5} shows the temporal evolution of some basic network properties, taken as the percolation point and with respect to their year-2002 values. One sees that the LD-PIN is continuously expanding in the years before the GFC'08 in terms of its number of nodes and edges $M$. Its edge density $\rho_p\equiv M/(N^2-N)$ is seen to track the aggregated market values of several over-the-counter-traded (OTC) financial derivative products, namely NOA-CDS and GMV-ELD \cite{bis_der}. Observing that these products somehow lag behind the network density and that their rates of change differ by about one order of magnitude, we use a non-linear short-term memory model (NLSMM,\ \cite{pin}) for their phenomenological description. 
For the total market value $V_{D}\left(t_n\right)$ (NOA or GMV) of some derivative $V_{D}$ at time $t_n$, with respect to a reference value $V_r=V_D\left(t_r\right)$, we write
\begin{equation}
V_{D}\left(t_n\right)\,=\,V_{r}\cdot a_r\,\left[\,\bar{\rho}^{\gamma_1}\left(t_n\right)\,+\,\bar{\rho}^{\gamma_2}\left(t_{n-1}\right)\,\right],
\quad\mbox{where}\quad
\bar{\rho}\left(t_n\right)\,\equiv \,\rho_{p}^{LD}(t_n)\,/\,\rho_{p}^{LD}(t_r)\,.
\label{eq:mm}
\end{equation}
Here, $t_{n-1}$ denotes the data point before $t_n$, which is the previous year for a 1-year period of the CPIS data \cite{cpis};
$\gamma_1$ and $\gamma_2$ are scaling exponents, where a value different from one implies non-linearity; $a_r$ is a scaling factor, accounting for the arbitrarily chosen reference value $V_r$. When using a least-square method to fit (\ref{eq:mm}) to NOA-CDS and GMV-ELD, we consider different lead/lag relations between the edge density and the market value of derivative products, where positive lead relations indicate that changes in the derivative market are a consequence of changes in the density of the LD-PIN. We find $\Delta t=6\,$months and $\Delta t=0\,$months for NOA-CDS and GMV-ELD, respectively. This suggests that a net-lead of $\rho_{p}$ with respect to the market value of financial derivative is the case, while the availability of larger data amounts would clarify these relations.\\
To use the NLSMM (\ref{eq:mm}) for the quantitative warning for potential financial crises, by indicating high levels of OTC-traded derivative products, one can set a maximal warning threshold $w_{th}$ which one deems still safe. A dynamical approach, which allows for market changes a certain product may be coupled to, is to set the warning threshold with respect to a macroeconomic reference variable (RV). If $V_{RV}\left(t\right)$ denotes the value of an RV at time $t$, we require
\begin{equation}
\label{eq:wth}
V_D\left(t\right)\,\leq\,w_{th}\left(t\right)\,=\,F^{D,RV}_{max}\cdot\,V_{RV}\left(t\right)\,,
\end{equation}
where $V_D$ is the market value of some derivative product (NOA or GMV), and $F^{D,RV}_{max}$ is a multiplier, setting the maximal proportion between $V_D$ and $V_{RV}$. This is a dynamic threshold, in the sense that, if $V_{RV}$ increases, $V_{D}$ is also allowed to increase. However, if $V_{RV}$ decreases, such that $V_D$ comes close to $w_{th}$, or if  $V_D$ exceeds the threshold, $V_D$ must be reduced by some mechanism. A regulatory approach, based on monetary incentives to enforce a certain ``equilibrium level'' $F^{D,RV}_{max}$, would be an RV-progressive transaction or holding tax on financial derivatives, which makes especially speculative trading unattractive, as soon as the level of derivative $D$ comes close to $F^{D,RV}_{max}$ \cite{pin}.\\
\begin{figure}
  \begin{mdframed}
  \begin{center}
    \includegraphics[width=1.\textwidth]{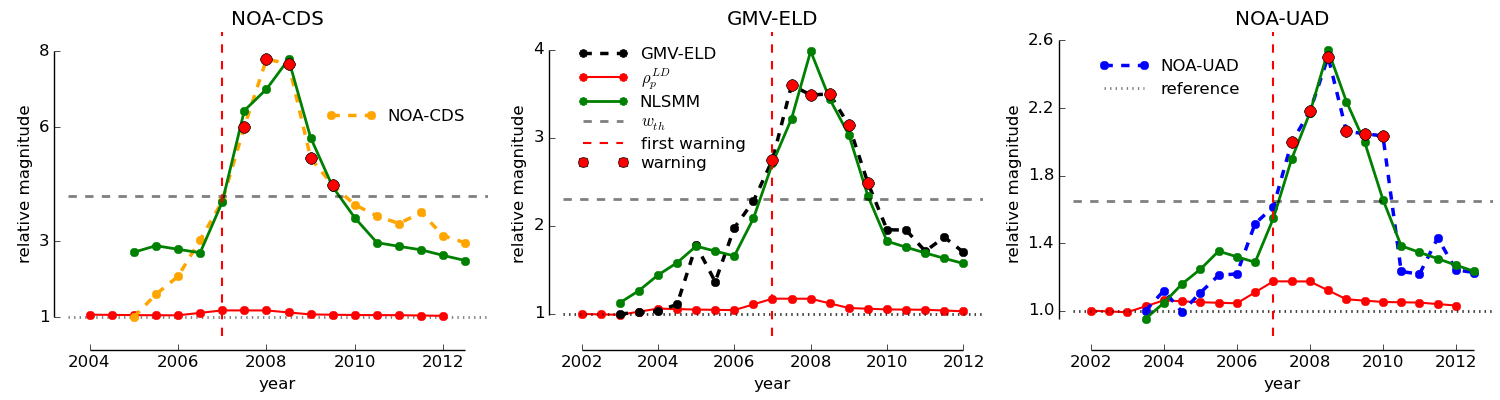}
  \end{center}
  \caption{\small{The \textbf{non-linear short-term memory model} (NLSMM) describing the proliferation of the notional outstanding amount of credit default swaps (NOA-CDS), gross-market value of equity-link derivatives (GMV-ELD) and notional outstanding amount of unallocated derivatives (NOA-UAD), \cite{bis_der,bis_tri,isda_3}. The edge density of the LD-PIN (red line) is taken as the model input, achieving an accurate description for all three products (green lines). Setting a warning threshold $w_{th}$ as a fraction of world-GDP (grey dashed line), consistent first warning signals (red dashed lines) are generated. The fit results for the given products are,
  NOA-CDS: $\gamma_1=11.0$, $\gamma_2=6.6$, $\Delta t=6\,$months, $w_{th}=0.56\,$world-GDP. 
  GMV-ELD: $\gamma_1=7.3$, $\gamma_2=8.0$,  $\Delta t=0$, $w_{th}=0.014\,$world-GDP.
  NOA-UAD: $\gamma_1=5.9$, $\gamma_2=6.0$,  $\Delta t=6\,$months, $w_{th}=0.75\,$world-GDP. Source:\ \cite{pin}.}}
\end{mdframed}
\label{fig:nlsmm}
\end{figure}
World-GDP is found to offer a suitable and widely used RV for the indication of the GFC'08, with respect to a variety of derivative products. \hyperref[fig:nlsmm]{Fig.\ 6} shows an application of the NLSMM (\ref{eq:mm}) for the description of NOA-CDS\footnote{the initial mismatch is explained through the late start of data taking in the end of 2004, which fell into a period of fast growth.}, GMV-ELD and unallocated derivatives (NOA-UAD), \cite{bis_der,bis_tri,isda_3}, which may be interpreted as a ``hidden variable'', which can be tracked well by the presented model. It can be seen that one mostly achieves a good description, especially around the years of the GFC'08. A warning threshold as a fraction of world-GDP $F^{D,GDP}_{max}$ has been set for all three products, where consistent first warning signals are generated at the beginning of 2007 \cite{pin}. Note that these results are valid for a large range of possible values of $F^{D,GDP}_{max}$, which hugely enhances the applicability of the presented mechanism.\\
To further evaluate the applicability of the proposed methodology, the NLSMM (\ref{eq:mm}) had been tested for the description of all main categories and first subcategories of OTC-traded financial derivative products, contained in \cite{bis_der}, and also some exchange-traded products. The detailed fit results are listed in \hyperref[tab:der]{Tab.\ 2}, where one can see that the NLSMM achieves good descriptions for CDS and ELD, as well as commodity-linked derivatives (CLD). The descriptions of CLD and ELD are generally better at lower values of the edge threshold, where the shaded area on the left side of \hyperref[fig:ld_props]{Fig.\ 5} indicates the allowable region for the NLSMM (\ref{eq:mm}).\\
The methodology presented in this section offers a prime example for how approaches from network science and computational economics can be coupled to generate new practical insights and applicable concepts. The recent financial crises might be seen as a testing ground, where such concepts can be probed to avoid future large-scale downturns.
%
%
\section{The G-20 Trade Network and Local Power}
\label{sec:g20}
\begin{figure}[h!]
  \begin{mdframed}
  \begin{center}
    \includegraphics[width=.35\textwidth]{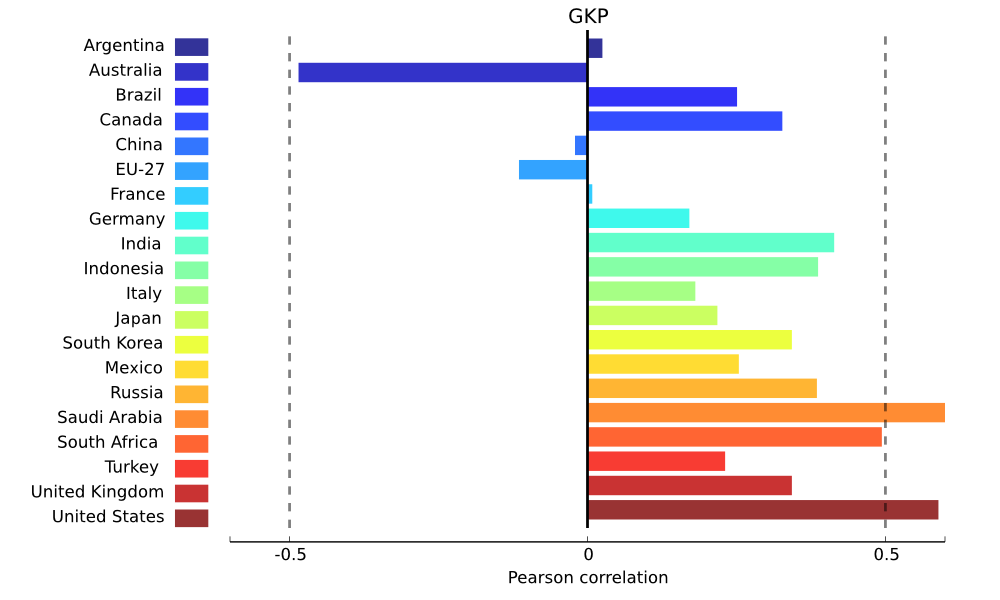}
    \includegraphics[width=.28\textwidth]{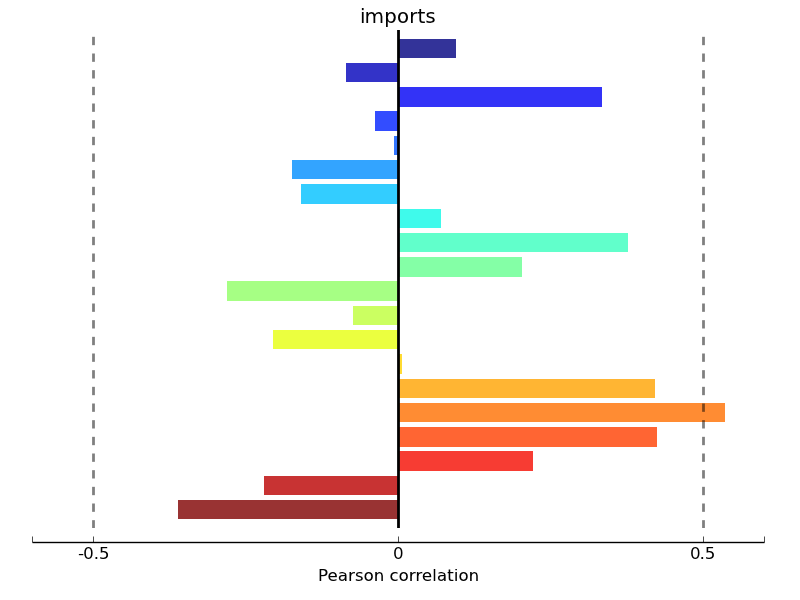}
    \includegraphics[width=.35\textwidth]{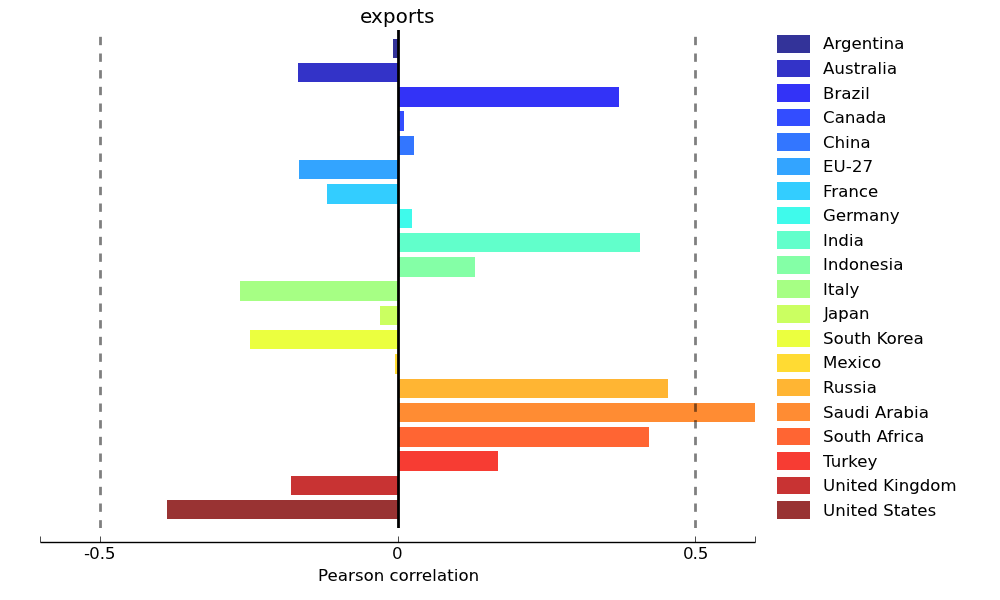}
  \end{center}
  \caption{\small{\textbf{Comparison of the correlation of the annual change of GDP} with the GKP (left) and the total amounts of imports (middle) and exports (right) for the group of G-20 economies between 1991-2011. While most countries show considerable correlations of their annual change in GDP with either of the three quantities, this relation is mostly positive only for the GKP, making it a powerful indicator for the projection of economic growth and development.}}
  \end{mdframed}
\label{fig:gkp-20}
\end{figure}

International trade is one of the main catalysts of globalisation and social and economic development. The volume of global trade grew by more than a factor of five, in nominal terms, over the past 20\ years\cite{comtrade}, thereby closely linking up many of the involved countries and forming a international web of trade. In this sense, the investigation of the relation of global trade and economic growth is highly suited for the application of techniques from network science. Trade flows between individual economic can be readily translated to form the edges of an \textit{international trade network} (ITN).\\
For the investigation of the ITN and for the demonstration of how simple network statistics can be used to project economic growth, we focus on the G-20 group of major economies\footnote{Argentina, Australia, Brazil, Canada, China, EU-27, France, Germany, India, Indonesia, Italy, Japan, South Korea, Mexico, Russia, Saudi Arabia, South Africa, Turkey, United Kingdom, United States.}, which make up for about $80\%$ of cross-border trade. The resulting trade network (ITN-20) is shown in \hyperref[fig:g20_wtw]{Fig.\ 1}, where the node size is set according to the GDP of that country (or a group of countries for the EU-27) and the thickness of arrows and their heads correspond to the volume of directed trade flows between two countries. One clearly sees the emergence of a global ``triangle of trade'' between Europe, the United States and China, and how individual countries connected to this macro-structure. As pointed out before, this illustrates how the network picture delivers a ``bird's eye view of complexity'' and how complex interrelations can be readily read out. \\
\begin{figure}
  \begin{mdframed}
  \begin{center}
    \includegraphics[width=.4\textwidth]{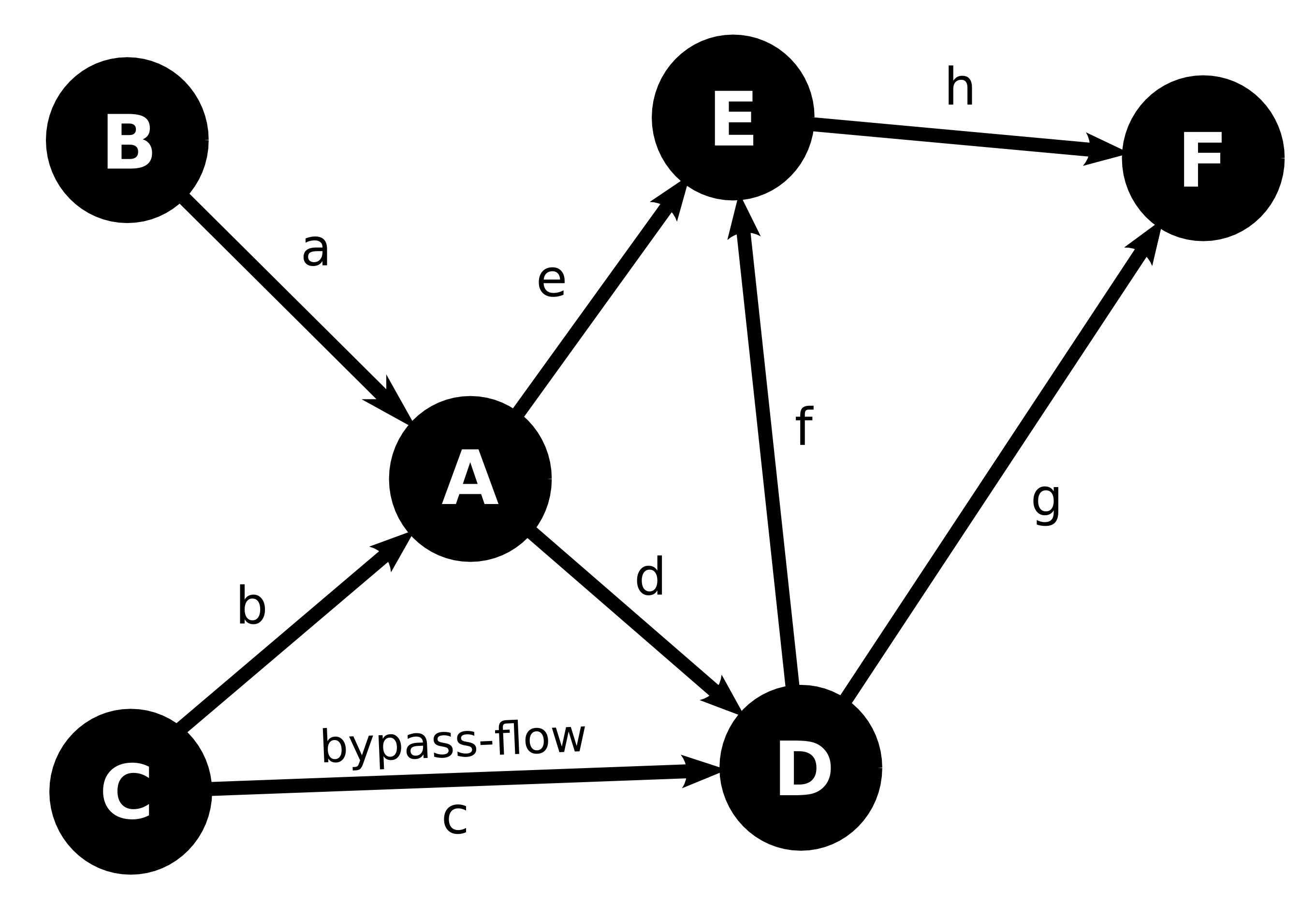}\hspace*{1cm}
    \includegraphics[width=.4\textwidth]{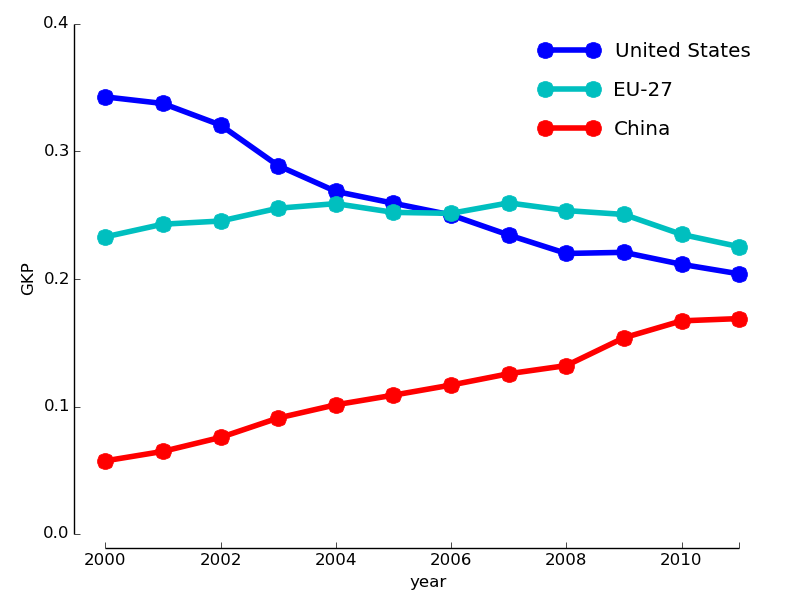}
  \end{center}
  \caption{\small{Left part: \textbf{Schematic working of the gate-keeping potential} (GKP), which is defined as the fraction of a node's nearest-neighbour flow, going through itself. Assuming unit edge weights (lower-case letters), the GKP of node A is $g\left(A\right)=\sqrt{(a+b)(e+d)}/\left(\left(\sqrt{(a+b)(e+d)}+c\right)\right)=2/3$.\\
Right part: \textbf{Temporal evolution of the GKP} for the global ``triangle of trade'', consisting of Europe, the United States and China (see \hyperref[fig:g20_wtw]{Fig.\ 1}). The GKP is interpreted as a measure for control within the ITN-20, by indicating a countries capability to funnel trade flows through itself. While the United States saw a decline of its GKP in recent years, China saw a steady rise. The projected convergence of all three scores may suggests an equilibration of the global trade system.\\[-.3cm]}}
\label{fig:gkp}
\end{mdframed}
\end{figure}
An instructive network statistic, which measures the amount of local control a node might exert over its neighbours is the \textit{gate-keeping potential} (GKP). It is defined as the fraction of a node's nearest-neighbour flow which goes through itself. Let $A$ be the flow matrix of a general network, as defined in the \hyperref[sec:intro]{Introduction}, then the GKP of node $i$ is given by
\begin{eqnarray}
g\left(i\right)\,&\equiv&\,\frac{\sqrt{(\mbox{in-flow})_i\,\times\,(\mbox{out-flow})_i}}{\sqrt{(\mbox{in-flow})_i
\,\times\,(\mbox{out-flow})_i}\,+\,\mbox{(bypassing flow}_i)} \\
&=&\,\frac{\sqrt{\sum_{k=0}^{N}\,A_{ki}\,\times\,\sum_{k=0}^{N}\,A_{ik}}}
{\sqrt{\sum_{k=0}^{N}\,A_{ki}\,\times\,\sum_{k=0}^{N}\,A_{ik}}\,+\,\left[\bar{A}^T\cdot A\cdot \bar{A}^T\right]_{ii}}\quad\in\quad [0,1]\,,\nonumber
\end{eqnarray}
where $\bar{A}^T$ is the transposed flow matrix, which only contains ones for each positive element of $A$. We set $g$ to zeros, if a node does not have any out- or in-flow, i.e.\ if it cannot act as a gate. The schematic working of the GKP is shown in \hyperref[fig:gkp]{Fig.\ 7 (left part)}. Considering node $A$, only down-stream nearest-neighbour connections enter the calculation, i.e.\ connection which either go into $A$, out of $A$ or from a node up-stream of $A$ to a node down-stream of $A$ (bypass-flow). Setting all flow values (lower case letters) to one, the GKP of $A$ evaluates to $g\left(A\right)=\sqrt{(a+b)(e+d)}/\left(\left(\sqrt{(a+b)(e+d)}+c\right)\right)=2/3$. The GKP can be seen as a measure for local power, since, assuming that some ``vital'' flow takes place on the network, a high GKP means that this node might exert large control (power) on the nodes up- or downstream of it. While a low GKP indicates that a node is not important for a flow process, since it can be easily bypassed.\\
Translating these concepts to the ITN-20, a high GKP indicates country's high capability to funnel trade flows through itself, making its individual trade partners dependent on a country's im- and exports. Note that the ITN-20 is densely connected, meaning that a high GKP score translates into global control over the whole network. The evolution of the GKP for the ``triangle of trade'' (Europe, the United States and China) between 2000-2011 is shown in \hyperref[fig:gkp]{Fig.\ 7 (right part)}. One see how the very dominant position of the United States had been reduced over this time, while China saw a steady rise of its score. The projected convergence of all three scores, may suggests an equilibration of the global trade system and a more balanced world-economy.\\
A country's import and export volumes are considered standard indicators for the strength of its economy, as is its annual change of GDP. As a consequence, one would expect that there is a clear relation between this set of major macroeconomic indicators. Looking at the middle and right panels of \hyperref[fig:gkp-20]{Fig.\ 8}, which show the Pearson product correlation coefficient between G-20 economies' annual change of GDP and import and export volumes, respectively, for the years 1991-2011, one sees that there is mostly a strong (or at least decent) connection between trade and change in GDP. However, these relations are not found to be uniform. For some countries they are strongly positive, as for Saudi Arabia and South Africa, while for other countries they are found to be negative, as for the United States\footnote{there are, of course, several additional factors which affect the change of GDP, which will not be considered further.}. Taking now the topological embedding of a country into account, in terms of its GKP, the above relations between trade and annual change of GDP turn out to be mostly positive. This is interpreted as making the GKP a useful indicator for the projection of economic development.\\
The major exception from this finding is Australia, which has a large negative correlation between its GKP and GDP change. This can be explained through the fact that Australia's major export goods are raw materials, which are produced within the country. Within the flow picture of the ITN-20, this means that Australia is not a ``gate'' for its major export goods, but rather a source. This example highlights once more, how the network approach to economics problems can be combined with conventional methods to obtain deeper insights into complex economic systems.
\clearpage
\section{Conclusion}
\label{sec:con}
In this report, we have presented different approaches of how the combination of data science (Big Data), network science and economics can be used to live up to the challenges of a globalised and interconnected world. In the \hyperref[sec:mlr]{first study}, a MLR-fit algorithm has been introduced to model the multiple links between individual components of balance of payment accounts of a group of countries, which encompasses the majority of global (macro)economic activity. The derived network model delivered a very accurate description of most indicators, while the built-in time shift between regressors and regressands led to good forecasts in the majority of cases. We have introduced the concept of an indicator's tracking centrality, which allowed for the identification of ``macroeconomic thermometers'', i.e.\ indicators which track multiple large indicators with a high precision. This novel methodology is supposed to provide powerful monitoring tools for complex processes of economic cross-border interactions.\\
In the \hyperref[sec:pin]{second example}, we have presented an early-warning mechanism for global financial crises, which combines network statistics with commonly-used quantities in macroeconomic analyses. It had been shown that the edge density of the long-term debt portfolio investment network can be used as the input variable for a non-linear phenomenological model to describe the proliferation of a wide range of financial derivative products, most prominently CDS. When coupling the model output to an economic standard variable, such as world-GDP, consistent warning signals are generated ahead of the recent global financial crisis of 2008, with respect to several derivative products. \\
In the \hyperref[sec:pin]{third case study}, we have investigated the international trade network of the G-20 group of major economies. We have demonstrated how the change of local topological properties can be used to describe global economic development, as well as the development of individual economies, during the last 10-20 years. For most countries, the gate-keeping potential, as a measure for local power, turns out to be a far better indicator for the projection of the annual change of GDP than the volume of imports or exports.\\
All of these methodologies are not meant to stand alone, but to be merged with established economic analysis tools, and to be fuelled by ever-increasing amounts of data to generate a network gain for a better understanding of a world which is ever more interconnected and interdependent.
\section*{Acknowledgements}
\addcontentsline{toc}{section}{Acknowledgements}
We would like to thank the European Commission FET Open Project FOC 255987 and FOC-INCO 297149 for financial support. The authors also thank the developers and supporters of the Python programming language, including IPython and NetworkX, which have been used throughout this study.
\addcontentsline{toc}{section}{References}

%
\clearpage
\begin{landscape}
\section*{Appendix: MLR-Fit Statistics}
\addcontentsline{toc}{section}{Appendix: MLR-Fit Statistics}
\label{sec:app}
\begin{center}
\small{
\begin{longtable}{|ll|rrcc||ll|rrcc|}
\hline
\multicolumn{1}{|l}{\cellcolor[gray]{0.9} country} & \multicolumn{1}{|l}{\cellcolor[gray]{0.9} Indicator} &
\multicolumn{1}{|l}{\cellcolor[gray]{0.9} ave. err.} & \multicolumn{1}{|l}{\cellcolor[gray]{0.9} fc. err.} &
\multicolumn{1}{|l}{\cellcolor[gray]{0.9} $\#$\ r'ors} & \multicolumn{1}{|l|}{\cellcolor[gray]{0.9} $\#$\ r'nds} &
\multicolumn{1}{||l}{\cellcolor[gray]{0.9} country} & \multicolumn{1}{|l}{\cellcolor[gray]{0.9} Indicator} &
\multicolumn{1}{|l}{\cellcolor[gray]{0.9} ave. err.} & \multicolumn{1}{|l}{\cellcolor[gray]{0.9} fc. err.} &
\multicolumn{1}{|l}{\cellcolor[gray]{0.9} $\#$\ r'ors} & \multicolumn{1}{|l|}{\cellcolor[gray]{0.9} $\#$\ r'nds} \\
\specialrule{.1em}{.05em}{.05em} 
Algeria & Imports & 3.7 & 0.2 & 1 & 0 & Algeria & Exports & 5.3 & 7.5 & 1 & 0\\
Algeria & FDI (in) & 1.0 & 10.8 & 3 & 0 & Algeria & FDI (out) & 2.9 & 22.7 & 2 & 1\\
Argentina & Imports & 2.2 & 5.3 & 2 & 3 & Argentina & Exports & 1.0 & 17.9 & 2 & 3\\
Argentina & FDI (in) & 4.1 & 10.9 & 1 & 2 & Argentina & FDI (out) & 3.1 & 9.5 & 2 & 1\\
Argentina & Equity (in) & 4.8 & 19.1 & 2 & 2 & Argentina & Equity (out) & 1.3 & 7.2 & 3 & 0\\
Argentina & Debt (in) & 4.6 & 14.6 & 2 & 3 & Argentina & Debt (out) & 3.1 & 8.4 & 2 & 0\\
Australia & Imports & 1.4 & 6.1 & 1 & 1 & Australia & Exports & 1.8 & 10.5 & 2 & 2\\
Australia & FDI (in) & 3.8 & 7.9 & 1 & 0 & Australia & FDI (out) & 0.5 & 13.6 & 3 & 0\\
Australia & Equity (in) & 2.7 & 10.6 & 1 & 3 & Australia & Equity (out) & 1.8 & 21.1 & 2 & 3\\
Australia & Debt (in) & 2.9 & 1.4 & 1 & 2 & Australia & Debt (out) & 5.7 & 6.2 & 2 & 0\\
Austria & Imports & 0.6 & 11.2 & 2 & 0 & Austria & Exports & 0.2 & 11.8 & 4 & 1\\
Austria & FDI (in) & 1.3 & 0.8 & 3 & 0 & Austria & FDI (out) & 1.3 & 3.3 & 3 & 0\\
Austria & Equity (in) & 8.3 & 10.3 & 1 & 1 & Austria & Equity (out) & 0.8 & 11.0 & 1 & 0\\
Austria & Debt (in) & 2.4 & 7.7 & 1 & 2 & Austria & Debt (out) & 0.7 & 0.8 & 2 & 0\\
Belgium & Imports & 0.7 & 8.7 & 1 & 0 & Belgium & Exports & 0.5 & 4.8 & 2 & 1\\
Belgium & FDI (in) & 7.6 & 59.6 & 1 & 5 & Belgium & FDI (out) & 5.4 & 6.7 & 1 & 6\\
Belgium & Equity (in) & 4.1 & 3.0 & 1 & 0 & Belgium & Equity (out) & 2.6 & 3.8 & 2 & 0\\
Belgium & Debt (in) & 1.6 & 24.0 & 1 & 1 & Belgium & Debt (out) & 2.5 & 1.8 & 1 & 0\\
Bermuda & FDI (out) & 8.3 & 90.5 & 2 & 0 & Bermuda & Equity (in) & 3.2 & 0.3 & 1 & 5\\
Bermuda & Equity (out) & 6.7 & 442.9 & 1 & 0 & Bermuda & Debt (in) & 3.8 & 5.8 & 1 & 0\\
Bermuda & Debt (out) & 5.1 & 47.2 & 1 & 2 & Brazil & Imports & 3.2 & 3.3 & 1 & 5\\
Brazil & Exports & 0.9 & 10.8 & 2 & 3 & Brazil & FDI (in) & 5.3 & 11.6 & 1 & 2\\
Brazil & FDI (out) & 3.1 & 2.0 & 2 & 0 & Brazil & Equity (out) & 7.1 & 5.3 & 2 & 1\\
Brazil & Debt (in) & 9.3 & 21.8 & 1 & 0 & Brazil & Debt (out) & 7.2 & 59.2 & 2 & 1\\
Canada & Imports & 0.4 & 1.6 & 3 & 0 & Canada & Exports & 1.0 & 0.4 & 2 & 0\\
Canada & FDI (in) & 0.7 & 7.4 & 2 & 0 & Canada & FDI (out) & 2.9 & 5.7 & 1 & 0\\
Canada & Equity (in) & 1.7 & 6.1 & 2 & 0 & Canada & Equity (out) & 0.4 & 4.0 & 3 & 3\\
Canada & Debt (in) & 3.2 & 12.7 & 1 & 0 & Canada & Debt (out) & 4.8 & 8.4 & 2 & 0\\
Cayman Islands & FDI (in) & 2.6 & 21.9 & 2 & 0 & Cayman Islands & FDI (out) & 2.3 & 1.1 & 2 & 1\\
Cayman Islands & Debt (in) & 1.1 & 20.7 & 2 & 3 & Cayman Islands & Debt (out) & 4.9 & 0.4 & 1 & 1\\
Chile & Imports & 2.3 & 11.3 & 2 & 0 & Chile & Exports & 4.8 & 0.8 & 1 & 8\\
Chile & FDI (in) & 1.5 & 0.3 & 2 & 3 & Chile & FDI (out) & 1.1 & 16.6 & 2 & 0\\
Chile & Equity (in) & 3.0 & 53.3 & 2 & 3 & Chile & Equity (out) & 9.9 & 16.0 & 2 & 1\\
Chile & Debt (in) & 4.2 & 7.3 & 2 & 2 & China & Imports & 1.6 & 10.8 & 2 & 4\\
China & Exports & 2.1 & 0.8 & 1 & 6 & China & FDI (in) & 3.4 & 0.7 & 2 & 0\\
China & FDI (out) & 3.9 & 19.4 & 3 & 4 & Colombia & Imports & 1.0 & 0.3 & 2 & 1\\
Colombia & Exports & 1.0 & 3.1 & 2 & 2 & Colombia & FDI (in) & 3.7 & 45.2 & 2 & 1\\
Colombia & FDI (out) & 4.4 & 21.6 & 2 & 4 & Colombia & Equity (out) & 3.0 & 17.4 & 3 & 7\\
Colombia & Debt (in) & 0.8 & 18.1 & 4 & 0 & Colombia & Debt (out) & 6.2 & 16.5 & 2 & 1\\
Czech Republic & Imports & 1.2 & 18.6 & 2 & 0 & Czech Republic & Exports & 0.6 & 6.4 & 3 & 1\\
Czech Republic & FDI (in) & 1.3 & 15.4 & 2 & 0 & Czech Republic & FDI (out) & 6.7 & 17.3 & 2 & 4\\
Czech Republic & Equity (in) & 2.3 & 14.3 & 2 & 1 & Czech Republic & Equity (out) & 3.2 & 15.6 & 3 & 3\\
\hline
\multicolumn{1}{|l}{\cellcolor[gray]{0.9} country} & \multicolumn{1}{|l}{\cellcolor[gray]{0.9} Indicator} &
\multicolumn{1}{|l}{\cellcolor[gray]{0.9} ave. err.} & \multicolumn{1}{|l}{\cellcolor[gray]{0.9} fc. err.} &
\multicolumn{1}{|l}{\cellcolor[gray]{0.9} $\#$\ r'ors} & \multicolumn{1}{|l|}{\cellcolor[gray]{0.9} $\#$\ r'nds} &
\multicolumn{1}{||l}{\cellcolor[gray]{0.9} country} & \multicolumn{1}{|l}{\cellcolor[gray]{0.9} Indicator} &
\multicolumn{1}{|l}{\cellcolor[gray]{0.9} ave. err.} & \multicolumn{1}{|l}{\cellcolor[gray]{0.9} fc. err.} &
\multicolumn{1}{|l}{\cellcolor[gray]{0.9} $\#$\ r'ors} & \multicolumn{1}{|l|}{\cellcolor[gray]{0.9} $\#$\ r'nds} \\
\specialrule{.1em}{.05em}{.05em} 
Czech Republic & Debt (in) & 6.5 & 29.7 & 2 & 1 & Czech Republic & Debt (out) & 2.4 & 11.7 & 2 & 2\\
Denmark & Imports & 0.5 & 2.9 & 2 & 1 & Denmark & Exports & 0.9 & 13.1 & 2 & 4\\
Denmark & FDI (in) & 1.7 & 41.3 & 2 & 1 & Denmark & FDI (out) & 2.2 & 11.6 & 2 & 0\\
Denmark & Equity (in) & 0.8 & 24.2 & 3 & 1 & Denmark & Equity (out) & 0.8 & 3.7 & 2 & 0\\
Denmark & Debt (in) & 2.1 & 17.3 & 2 & 1 & Denmark & Debt (out) & 2.0 & 5.7 & 2 & 0\\
Egypt & Imports & 7.8 & 4.0 & 1 & 3 & Egypt & Exports & 2.5 & 15.8 & 2 & 1\\
Egypt & FDI (in) & 3.1 & 21.0 & 2 & 1 & Egypt & FDI (out) & 2.0 & 5.7 & 4 & 1\\
Finland & Imports & 1.2 & 7.1 & 2 & 2 & Finland & Exports & 1.0 & 2.5 & 3 & 1\\
Finland & FDI (in) & 0.7 & 11.3 & 3 & 0 & Finland & FDI (out) & 0.6 & 2.7 & 3 & 0\\
Finland & Equity (in) & 1.6 & 42.9 & 3 & 10 & Finland & Equity (out) & 2.0 & 8.9 & 2 & 4\\
Finland & Debt (in) & 0.9 & 14.3 & 3 & 3 & Finland & Debt (out) & 1.7 & 4.4 & 1 & 1\\
France & Imports & 1.0 & 3.9 & 1 & 0 & France & Exports & 0.8 & 5.1 & 2 & 0\\
France & FDI (in) & 2.3 & 7.3 & 2 & 5 & France & FDI (out) & 2.5 & 0.2 & 2 & 1\\
France & Equity (in) & 0.8 & 15.5 & 2 & 0 & France & Equity (out) & 0.4 & 15.5 & 4 & 5\\
France & Debt (in) & 1.6 & 7.9 & 1 & 1 & France & Debt (out) & 1.5 & 15.7 & 2 & 0\\
Germany & Imports & 0.6 & 8.4 & 2 & 0 & Germany & Exports & 0.8 & 6.7 & 1 & 0\\
Germany & FDI (in) & 3.0 & 2.1 & 1 & 0 & Germany & FDI (out) & 1.3 & 2.6 & 2 & 1\\
Germany & Equity (in) & 2.0 & 15.6 & 2 & 0 & Germany & Equity (out) & 3.4 & 11.0 & 1 & 2\\
Germany & Debt (in) & 1.0 & 7.0 & 2 & 1 & Germany & Debt (out) & 0.7 & 0.8 & 2 & 1\\
Greece & Imports & 3.4 & 1.1 & 1 & 4 & Greece & Exports & 0.6 & 9.1 & 3 & 2\\
Greece & FDI (in) & 5.2 & 38.4 & 1 & 1 & Greece & FDI (out) & 2.4 & 3.6 & 2 & 1\\
Greece & Equity (out) & 4.3 & 17.2 & 2 & 0 & Greece & Debt (in) & 1.6 & 239.2 & 2 & 2\\
Greece & Debt (out) & 5.5 & 13.0 & 2 & 1 & Guernsey & Equity (in) & 2.8 & 3.8 & 2 & 0\\
Guernsey & Equity (out) & 3.2 & 4.8 & 2 & 1 & Guernsey & Debt (in) & 7.0 & 10.4 & 1 & 1\\
Guernsey & Debt (out) & 0.9 & 7.3 & 3 & 4 & Hong Kong & Imports & 2.4 & 7.4 & 1 & 0\\
Hong Kong & Exports & 1.2 & 6.2 & 2 & 0 & Hong Kong & FDI (in) & 2.3 & 28.6 & 2 & 4\\
Hong Kong & FDI (out) & 1.7 & 19.6 & 2 & 2 & Hong Kong & Equity (in) & 4.7 & 53.6 & 1 & 2\\
Hong Kong & Equity (out) & 1.9 & 36.3 & 3 & 0 & Hong Kong & Debt (in) & 7.7 & 45.3 & 2 & 4\\
Hong Kong & Debt (out) & 0.9 & 26.5 & 2 & 2 & Hungary & Imports & 1.9 & 6.4 & 1 & 1\\
Hungary & Exports & 0.9 & 12.5 & 2 & 3 & Hungary & FDI (in) & 2.8 & 16.2 & 1 & 0\\
Hungary & FDI (out) & 2.4 & 6.9 & 2 & 2 & Hungary & Equity (in) & 6.6 & 52.6 & 2 & 4\\
Hungary & Equity (out) & 5.6 & 56.8 & 3 & 6 & Hungary & Debt (in) & 4.6 & 5.1 & 1 & 0\\
India & Imports & 1.7 & 5.4 & 3 & 1 & India & Exports & 1.1 & 10.3 & 3 & 3\\
India & FDI (in) & 3.8 & 0.9 & 1 & 4 & India & FDI (out) & 8.7 & 23.0 & 1 & 2\\
Indonesia & Imports & 2.9 & 13.5 & 2 & 6 & Indonesia & Exports & 1.1 & 2.6 & 2 & 0\\
Indonesia & FDI (in) & 4.7 & 3.3 & 3 & 4 & Indonesia & Equity (in) & 3.9 & 4.0 & 2 & 1\\
Indonesia & Equity (out) & 9.2 & 13.4 & 3 & 0 & Indonesia & Debt (in) & 5.3 & 12.1 & 2 & 1\\
Indonesia & Debt (out) & 9.3 & 9.1 & 3 & 4 & Ireland & Imports & 1.2 & 22.6 & 2 & 0\\
Ireland & Exports & 2.2 & 7.5 & 2 & 2 & Ireland & FDI (in) & 2.6 & 26.4 & 2 & 2\\
Ireland & FDI (out) & 2.5 & 15.9 & 3 & 3 & Ireland & Equity (in) & 1.3 & 1.6 & 2 & 1\\
Ireland & Equity (out) & 0.6 & 3.1 & 3 & 0 & Ireland & Debt (in) & 2.7 & 3.5 & 2 & 3\\
Ireland & Debt (out) & 2.8 & 13.8 & 2 & 3 & Israel & Imports & 1.2 & 1.2 & 2 & 2\\
Israel & Exports & 0.3 & 7.0 & 2 & 1 & Israel & FDI (in) & 2.2 & 9.3 & 2 & 2\\
Israel & FDI (out) & 2.9 & 5.4 & 2 & 2 & Israel & Equity (in) & 3.3 & 57.8 & 2 & 0\\
\hline
\multicolumn{1}{|l}{\cellcolor[gray]{0.9} country} & \multicolumn{1}{|l}{\cellcolor[gray]{0.9} Indicator} &
\multicolumn{1}{|l}{\cellcolor[gray]{0.9} ave. err.} & \multicolumn{1}{|l}{\cellcolor[gray]{0.9} fc. err.} &
\multicolumn{1}{|l}{\cellcolor[gray]{0.9} $\#$\ r'ors} & \multicolumn{1}{|l|}{\cellcolor[gray]{0.9} $\#$\ r'nds} &
\multicolumn{1}{||l}{\cellcolor[gray]{0.9} country} & \multicolumn{1}{|l}{\cellcolor[gray]{0.9} Indicator} &
\multicolumn{1}{|l}{\cellcolor[gray]{0.9} ave. err.} & \multicolumn{1}{|l}{\cellcolor[gray]{0.9} fc. err.} &
\multicolumn{1}{|l}{\cellcolor[gray]{0.9} $\#$\ r'ors} & \multicolumn{1}{|l|}{\cellcolor[gray]{0.9} $\#$\ r'nds} \\
\specialrule{.1em}{.05em}{.05em} 
Israel & Equity (out) & 5.9 & 5.1 & 2 & 2 & Israel & Debt (in) & 3.1 & 26.1 & 1 & 0\\
Israel & Debt (out) & 1.7 & 5.4 & 2 & 0 & Italy & Imports & 0.5 & 7.6 & 2 & 1\\
Italy & Exports & 0.5 & 0.9 & 3 & 1 & Italy & FDI (in) & 0.6 & 18.7 & 3 & 1\\
Italy & FDI (out) & 1.2 & 0.7 & 3 & 1 & Italy & Equity (in) & 3.8 & 23.6 & 1 & 1\\
Italy & Equity (out) & 1.2 & 12.2 & 2 & 0 & Italy & Debt (in) & 1.2 & 10.6 & 2 & 2\\
Italy & Debt (out) & 1.9 & 11.6 & 2 & 0 & Japan & Imports & 1.3 & 2.2 & 1 & 3\\
Japan & Exports & 1.3 & 0.4 & 1 & 3 & Japan & FDI (in) & 8.1 & 53.7 & 1 & 0\\
Japan & FDI (out) & 5.9 & 16.7 & 1 & 0 & Japan & Equity (in) & 6.2 & 1.9 & 1 & 3\\
Japan & Equity (out) & 1.9 & 1.5 & 2 & 1 & Japan & Debt (in) & 6.0 & 12.5 & 1 & 0\\
Japan & Debt (out) & 3.8 & 7.6 & 1 & 1 & Jersey & Equity (in) & 9.3 & 41.2 & 1 & 3\\
Jersey & Equity (out) & 3.6 & 11.1 & 1 & 1 & Jersey & Debt (in) & 8.5 & 7.4 & 1 & 2\\
Jersey & Debt (out) & 8.1 & 82.1 & 1 & 7 & Kazakhstan & Imports & 4.5 & 11.2 & 1 & 1\\
Kazakhstan & Exports & 6.9 & 3.4 & 1 & 1 & Kazakhstan & FDI (in) & 2.6 & 4.9 & 2 & 2\\
Kazakhstan & Equity (out) & 5.2 & 20.9 & 2 & 0 & Kazakhstan & Debt (in) & 3.6 & 94.8 & 4 & 6\\
Kazakhstan & Debt (out) & 3.3 & 23.7 & 2 & 4 & Luxembourg & Imports & 0.4 & 16.6 & 3 & 1\\
Luxembourg & Exports & 0.5 & 8.8 & 3 & 2 & Luxembourg & FDI (in) & 4.8 & 7.7 & 1 & 1\\
Luxembourg & FDI (out) & 6.4 & 41.6 & 1 & 1 & Luxembourg & Equity (in) & 2.2 & 3.7 & 1 & 0\\
Luxembourg & Equity (out) & 1.5 & 6.0 & 1 & 0 & Luxembourg & Debt (in) & 1.4 & 2.5 & 2 & 4\\
Luxembourg & Debt (out) & 0.7 & 0.8 & 2 & 0 & Malaysia & Imports & 0.3 & 6.0 & 3 & 1\\
Malaysia & Exports & 0.2 & 0.8 & 3 & 1 & Malaysia & FDI (in) & 1.2 & 4.8 & 2 & 4\\
Malaysia & FDI (out) & 5.3 & 9.0 & 2 & 2 & Malaysia & Equity (in) & 4.5 & 3.0 & 2 & 3\\
Malaysia & Equity (out) & 8.4 & 23.6 & 3 & 0 & Malaysia & Debt (in) & 1.3 & 55.1 & 3 & 0\\
Malaysia & Debt (out) & 2.2 & 8.3 & 3 & 4 & Mexico & Imports & 1.1 & 3.1 & 1 & 0\\
Mexico & Exports & 0.9 & 15.2 & 2 & 2 & Mexico & FDI (in) & 3.2 & 41.2 & 1 & 2\\
Mexico & FDI (out) & 2.9 & 4.1 & 2 & 3 & Netherlands & Imports & 0.9 & 0.2 & 1 & 2\\
Netherlands & Exports & 0.3 & 5.1 & 3 & 1 & Netherlands & FDI (in) & 1.1 & 1.5 & 2 & 0\\
Netherlands & FDI (out) & 2.3 & 18.2 & 1 & 1 & Netherlands & Equity (in) & 1.4 & 36.2 & 2 & 2\\
Netherlands & Equity (out) & 2.2 & 12.0 & 2 & 1 & Netherlands & Debt (in) & 1.2 & 1.5 & 2 & 0\\
Netherlands & Debt (out) & 0.5 & 8.0 & 3 & 0 & New Zealand & Imports & 1.7 & 3.6 & 1 & 2\\
New Zealand & Exports & 1.0 & 9.6 & 2 & 3 & New Zealand & FDI (in) & 2.1 & 14.6 & 1 & 2\\
New Zealand & FDI (out) & 1.6 & 1.1 & 3 & 1 & New Zealand & Equity (in) & 0.9 & 9.4 & 3 & 1\\
New Zealand & Equity (out) & 5.8 & 21.3 & 1 & 1 & Nigeria & FDI (in) & 4.8 & 0.6 & 1 & 3\\
Norway & Imports & 1.5 & 3.7 & 1 & 1 & Norway & Exports & 3.3 & 5.3 & 1 & 0\\
Norway & FDI (in) & 2.2 & 16.8 & 2 & 0 & Norway & FDI (out) & 2.7 & 8.3 & 1 & 2\\
Norway & Equity (in) & 2.7 & 39.1 & 3 & 2 & Norway & Equity (out) & 2.9 & 4.3 & 2 & 3\\
Norway & Debt (in) & 2.8 & 2.4 & 2 & 2 & Norway & Debt (out) & 3.8 & 8.4 & 1 & 0\\
Philippines & Imports & 1.1 & 25.6 & 2 & 0 & Philippines & Exports & 3.0 & 12.4 & 1 & 0\\
Philippines & FDI (in) & 5.3 & 4.3 & 1 & 0 & Philippines & FDI (out) & 2.1 & 1.5 & 3 & 0\\
Philippines & Equity (in) & 3.0 & 18.8 & 3 & 2 & Philippines & Debt (in) & 8.5 & 10.7 & 1 & 3\\
Philippines & Debt (out) & 1.6 & 30.3 & 3 & 0 & Poland & Imports & 1.0 & 9.2 & 3 & 1\\
Poland & Exports & 1.3 & 19.4 & 2 & 3 & Poland & FDI (in) & 2.8 & 11.3 & 1 & 1\\
Poland & Equity (in) & 6.2 & 44.3 & 1 & 5 & Poland & Debt (in) & 2.9 & 23.4 & 2 & 2\\
Portugal & Imports & 2.7 & 2.0 & 2 & 0 & Portugal & Exports & 2.1 & 14.5 & 2 & 3\\
Portugal & FDI (in) & 0.9 & 49.9 & 3 & 1 & Portugal & FDI (out) & 3.4 & 9.7 & 1 & 2\\
\hline
\multicolumn{1}{|l}{\cellcolor[gray]{0.9} country} & \multicolumn{1}{|l}{\cellcolor[gray]{0.9} Indicator} &
\multicolumn{1}{|l}{\cellcolor[gray]{0.9} ave. err.} & \multicolumn{1}{|l}{\cellcolor[gray]{0.9} fc. err.} &
\multicolumn{1}{|l}{\cellcolor[gray]{0.9} $\#$\ r'ors} & \multicolumn{1}{|l|}{\cellcolor[gray]{0.9} $\#$\ r'nds} &
\multicolumn{1}{||l}{\cellcolor[gray]{0.9} country} & \multicolumn{1}{|l}{\cellcolor[gray]{0.9} Indicator} &
\multicolumn{1}{|l}{\cellcolor[gray]{0.9} ave. err.} & \multicolumn{1}{|l}{\cellcolor[gray]{0.9} fc. err.} &
\multicolumn{1}{|l}{\cellcolor[gray]{0.9} $\#$\ r'ors} & \multicolumn{1}{|l|}{\cellcolor[gray]{0.9} $\#$\ r'nds} \\
\specialrule{.1em}{.05em}{.05em} 
Portugal & Equity (in) & 3.7 & 4.6 & 3 & 0 & Portugal & Equity (out) & 1.5 & 3.8 & 3 & 0\\
Portugal & Debt (in) & 1.7 & 30.4 & 3 & 1 & Portugal & Debt (out) & 3.2 & 24.5 & 1 & 1\\
Romania & Imports & 2.2 & 26.7 & 2 & 5 & Romania & Exports & 0.6 & 10.3 & 3 & 3\\
Romania & FDI (in) & 5.8 & 48.5 & 1 & 3 & Romania & FDI (out) & 4.6 & 65.5 & 2 & 3\\
Romania & Debt (in) & 1.2 & 5.8 & 3 & 3 & Russian Federation & Imports & 6.2 & 33.8 & 1 & 2\\
Russian Federation & Exports & 2.1 & 1.5 & 2 & 0 & Russian Federation & FDI (out) & 8.1 & 7.3 & 1 & 1\\
Russian Federation & Debt (in) & 8.6 & 71.4 & 2 & 5 & Russian Federation & Debt (out) & 5.1 & 28.3 & 3 & 5\\
Saudi Arabia & FDI (in) & 9.1 & 4.7 & 1 & 1 & Saudi Arabia & FDI (out) & 5.1 & 9.1 & 1 & 0\\
Singapore & Imports & 0.2 & 4.6 & 3 & 1 & Singapore & Exports & 0.6 & 4.3 & 2 & 1\\
Singapore & FDI (in) & 1.2 & 0.8 & 2 & 1 & Singapore & FDI (out) & 0.5 & 5.2 & 2 & 0\\
Singapore & Equity (in) & 5.2 & 7.5 & 1 & 2 & Singapore & Equity (out) & 1.0 & 8.0 & 3 & 0\\
Singapore & Debt (in) & 6.0 & 33.4 & 2 & 3 & Singapore & Debt (out) & 4.1 & 8.9 & 1 & 0\\
Slovak Republic & Imports & 1.2 & 19.8 & 2 & 1 & Slovak Republic & Exports & 0.8 & 13.0 & 2 & 1\\
Slovak Republic & FDI (in) & 4.2 & 8.1 & 1 & 2 & Slovak Republic & Equity (out) & 6.8 & 6.7 & 2 & 1\\
Slovak Republic & Debt (in) & 8.6 & 0.9 & 1 & 2 & South Africa & Imports & 0.4 & 7.6 & 3 & 5\\
South Africa & Exports & 2.3 & 11.8 & 1 & 0 & South Africa & FDI (in) & 6.6 & 3.1 & 1 & 2\\
South Africa & FDI (out) & 2.3 & 28.8 & 2 & 0 & South Africa & Equity (in) & 6.1 & 12.0 & 1 & 0\\
South Africa & Equity (out) & 7.1 & 4.3 & 1 & 0 & South Africa & Debt (in) & 5.6 & 47.2 & 2 & 2\\
South Africa & Debt (out) & 3.3 & 66.3 & 2 & 1 & South Korea & Imports & 0.8 & 10.3 & 2 & 1\\
South Korea & Exports & 1.3 & 3.6 & 1 & 8 & South Korea & FDI (in) & 5.1 & 1.6 & 1 & 2\\
South Korea & FDI (out) & 5.3 & 16.0 & 1 & 2 & South Korea & Equity (in) & 3.6 & 97.8 & 2 & 0\\
South Korea & Debt (in) & 3.5 & 5.5 & 2 & 7 & Spain & Imports & 1.0 & 2.2 & 2 & 2\\
Spain & Exports & 0.6 & 7.3 & 2 & 0 & Spain & FDI (in) & 3.5 & 6.5 & 1 & 0\\
Spain & FDI (out) & 1.6 & 3.2 & 1 & 0 & Spain & Equity (in) & 4.4 & 3.1 & 1 & 3\\
Spain & Equity (out) & 1.3 & 6.4 & 3 & 1 & Spain & Debt (in) & 2.9 & 12.5 & 2 & 3\\
Spain & Debt (out) & 2.4 & 43.2 & 2 & 3 & Sweden & Imports & 0.8 & 13.0 & 2 & 0\\
Sweden & Exports & 0.6 & 3.0 & 2 & 0 & Sweden & FDI (in) & 1.9 & 9.3 & 2 & 0\\
Sweden & FDI (out) & 2.6 & 2.3 & 1 & 0 & Sweden & Equity (in) & 1.8 & 9.0 & 2 & 0\\
Sweden & Equity (out) & 0.8 & 7.6 & 3 & 0 & Sweden & Debt (in) & 2.9 & 7.9 & 1 & 0\\
Sweden & Debt (out) & 2.9 & 7.8 & 1 & 1 & Switzerland & Imports & 1.8 & 23.2 & 1 & 1\\
Switzerland & Exports & 0.9 & 5.7 & 2 & 2 & Switzerland & FDI (in) & 6.1 & 3.6 & 1 & 0\\
Switzerland & FDI (out) & 0.8 & 9.5 & 3 & 1 & Switzerland & Equity (in) & 3.0 & 2.7 & 1 & 1\\
Switzerland & Equity (out) & 1.2 & 2.8 & 2 & 3 & Switzerland & Debt (in) & 2.6 & 3.2 & 2 & 3\\
Switzerland & Debt (out) & 2.0 & 2.6 & 2 & 5 & Taiwan & FDI (in) & 3.6 & 20.0 & 1 & 1\\
Taiwan & FDI (out) & 2.4 & 8.2 & 2 & 0 & Thailand & Imports & 2.1 & 6.3 & 1 & 1\\
Thailand & Exports & 0.4 & 8.5 & 2 & 1 & Thailand & FDI (in) & 2.7 & 2.7 & 1 & 1\\
Thailand & FDI (out) & 5.4 & 36.3 & 1 & 2 & Thailand & Equity (in) & 7.9 & 4.0 & 1 & 0\\
Thailand & Equity (out) & 3.1 & 24.8 & 3 & 4 & Thailand & Debt (in) & 2.0 & 14.0 & 3 & 0\\
Thailand & Debt (out) & 4.5 & 36.1 & 3 & 2 & Turkey & Imports & 0.2 & 5.5 & 4 & 1\\
Turkey & Exports & 1.8 & 11.5 & 1 & 0 & Turkey & FDI (out) & 5.0 & 12.8 & 1 & 1\\
Turkey & Equity (in) & 2.1 & 145.3 & 3 & 1 & Turkey & Equity (out) & 2.6 & 6.4 & 3 & 4\\
Turkey & Debt (in) & 4.0 & 13.3 & 2 & 2 & Ukraine & Imports & 4.3 & 7.5 & 2 & 0\\
Ukraine & Exports & 2.9 & 5.1 & 2 & 2 & Ukraine & FDI (in) & 2.8 & 13.8 & 2 & 0\\
Ukraine & Equity (out) & 6.4 & 26.8 & 2 & 1 & United Arab Emirates & FDI (in) & 8.1 & 30.0 & 2 & 1\\
\hline
\multicolumn{1}{|l}{\cellcolor[gray]{0.9} country} & \multicolumn{1}{|l}{\cellcolor[gray]{0.9} Indicator} &
\multicolumn{1}{|l}{\cellcolor[gray]{0.9} ave. err.} & \multicolumn{1}{|l}{\cellcolor[gray]{0.9} fc. err.} &
\multicolumn{1}{|l}{\cellcolor[gray]{0.9} $\#$\ r'ors} & \multicolumn{1}{|l|}{\cellcolor[gray]{0.9} $\#$\ r'nds} &
\multicolumn{1}{||l}{\cellcolor[gray]{0.9} country} & \multicolumn{1}{|l}{\cellcolor[gray]{0.9} Indicator} &
\multicolumn{1}{|l}{\cellcolor[gray]{0.9} ave. err.} & \multicolumn{1}{|l}{\cellcolor[gray]{0.9} fc. err.} &
\multicolumn{1}{|l}{\cellcolor[gray]{0.9} $\#$\ r'ors} & \multicolumn{1}{|l|}{\cellcolor[gray]{0.9} $\#$\ r'nds} \\
\specialrule{.1em}{.05em}{.05em} 
United Arab Emirates & FDI (out) & 5.8 & 8.0 & 2 & 1 & United Kingdom & Imports & 0.2 & 17.0 & 3 & 0\\
United Kingdom & Exports & 1.2 & 2.6 & 2 & 0 & United Kingdom & FDI (in) & 0.9 & 3.2 & 2 & 2\\
United Kingdom & FDI (out) & 3.5 & 4.1 & 1 & 1 & United Kingdom & Equity (in) & 0.9 & 4.9 & 2 & 1\\
United Kingdom & Equity (out) & 2.7 & 9.2 & 1 & 3 & United Kingdom & Debt (in) & 0.8 & 0.7 & 2 & 3\\
United Kingdom & Debt (out) & 3.4 & 6.9 & 1 & 0 & United States & Imports & 0.8 & 6.1 & 2 & 2\\
United States & Exports & 0.5 & 6.3 & 2 & 1 & United States & FDI (in) & 1.5 & 12.3 & 2 & 0\\
United States & FDI (out) & 4.4 & 11.0 & 1 & 0 & United States & Equity (in) & 0.6 & 7.7 & 3 & 2\\
United States & Equity (out) & 1.4 & 6.5 & 1 & 1 & United States & Debt (in) & 1.2 & 6.5 & 1 & 0\\
United States & Debt (out) & 2.4 & 12.5 & 1 & 1 & Venezuela & FDI (in) & 3.4 & 4.4 & 2 & 2\\
Venezuela & FDI (out) & 4.3 & 20.8 & 1 & 1 & Venezuela & Debt (in) & 2.1 & 18.1 & 2 & 0\\
Venezuela & Debt (out) & 4.3 & 31.6 & 2 & 3 & Viet Nam & FDI (in) & 2.7 & 7.5 & 2 & 1\\
Brit. Virgin Islands & FDI (in) & 4.0 & 8.3 & 3 & 1 & Brit. Virgin Islands & FDI (out) & 4.1 & 8.6 & 1 & 0\\
\hline
\caption{Summary of MLR-fit statistics for all 372 indicators with s positive fit result, according to the linear model (\ref{eq:mlr}) and the set statistical requirements on the maximal admissible error, significance and collinearity. The columns for each country-indicator pair show the achieved time-averaged error, the forecasting error for the year-2012 value, the number of regressors (in-degree) and the number an indicator is regressor itself (the number of its regressands, out-degree), respectively. The average fit error for all indicators is $2.9\%$ and the median forecasting error is $8.5\%$. We give the median error here, since there are several outliers, which make the mean forecasting error of $15.9\%$ unrepresentative.}
\label{tab:mlr_stats}
\end{longtable}
}
\end{center}
\end{landscape}
\section*{Appendix: NLSMM Description of Financial Derivatives}
\addcontentsline{toc}{section}{Appendix: NLSMM Description of Financial Derivatives}
\begin{table}[ht!]
\begin{center}
\small{
\renewcommand{\arraystretch}{1}
\begin{tabular}{|l|l|l||r|r|r|r|r|r|r|l|}
\hline
\multicolumn{3}{|l||}{ \cellcolor[gray]{0.9} name} & \multicolumn{1}{c|}{\cellcolor[gray]{0.9} $f_{GDP}^{\mathrm{world}}$} 
 & \multicolumn{1}{c|}{\cellcolor[gray]{0.9} $a_r$ } & \multicolumn{1}{c|}{\cellcolor[gray]{0.9} $\gamma_1$} 
 & \multicolumn{1}{c|}{\cellcolor[gray]{0.9} $\gamma_2$ } & \multicolumn{1}{c|}{\cellcolor[gray]{0.9} $m$} 
 & \multicolumn{1}{c|}{\cellcolor[gray]{0.9} $\Delta\,t$}  & \multicolumn{1}{c|}{\cellcolor[gray]{0.9} $p_r$} 
 & \multicolumn{1}{c|}{\cellcolor[gray]{0.9}  decision} \\
\specialrule{.1em}{.05em}{.05em}
\multirow{21}{.7cm}{NOA} &
\multicolumn{2}{l||}{\textit{total}} & 10.939 & 0.7 & 4.1 & 3.3 & 0.8 & 12 & 0.7 & no\\
\cline{2-11}
 & \multirow{3}{.7cm}{CDS} & \textit{total} & 0.934 & 0.9 & 11.0 & 6.6 & 0.6 & 6 & 0.92 & \textbf{yes}\\
 &  & SNI & 0.543 & 0.7 & 9.6 & 6.9 & 0.7 & 6 & 0.93 & \textbf{yes}\\
 &  & MNI & 0.39 & 1.6 & 12.8 & 6.0 & 0.5 & 6 & 0.9 & \textbf{yes}\\
\cline{2-11}
 & \multirow{3}{.7cm}{FXD} & \textit{total} & 1.024 & 0.6 & 3.4 & 2.6 & 0.8 & 6 & 0.64 & no\\
 &  & F$\&$S & 0.52 & 0.6 & -1.3 & 5.7 & - & -6 & 0.8 & no\\
 &  & SWP & 0.265 & 0.8 & 0.1 & 2.6 & 17.8 & 12 & 0.31 & no\\
 &  & OPT & 0.239 & 0.6 & 5.7 & 2.1 & 0.4 & 6 & 0.89 & maybe\\
\cline{2-11}
 & \multirow{3}{.7cm}{IRD} & \textit{total} & 7.454 & 0.8 & 2.7 & 3.8 & 1.4 & 12 & 0.62 & no\\
 &  & FWD & 0.64 & 1.1 & -4.1 & 6.3 & - & 12 & 0.53 & no\\
 &  & SWP & 5.803 & 0.8 & 2.8 & 3.7 & 1.3 & 12 & 0.62 & no\\
 &  & OPT & 1.011 & 0.7 & 5.6 & 3.2 & 0.6 & 6 & 0.85 & maybe\\
\cline{2-11}
 & \multirow{3}{.7cm}{ELD} & \textit{total} & 0.166 & 0.7 & 4.6 & 7.0 & 1.5 & -6 & 0.92 & \textbf{yes}\\
 &  & F$\&$S & 0.043 & 0.8 & 8.0 & 2.4 & 0.3 & 6 & 0.84 & no\\
 &  & OPT & 0.122 & 0.6 & 4.7 & 6.2 & 1.3 & -6 & 0.9 & \textbf{yes}\\
\cline{2-11}
 & \multirow{3}{.7cm}{CLD} & \textit{total} & 0.215 & 0.7 & 9.8 & 14.1 & 1.4 & -6 & 0.87 & maybe\\
 &  & GLD & 0.011 & 0.4 & 5.5 & -0.7 & - & 0 & 0.77 & no\\
 &  & OTH & 0.205 & 1.0 & 10.6 & 15.0 & 1.4 & -6 & 0.87 & maybe\\
 &  & F$\&$S  & 0.123 & 0.6 & 15.4 & 8.0 & 0.5 & 6 & 0.89 & maybe\\
 &  & OPT & 0.082 & 0.9 & 15.0 & 13.4 & 0.9 & -6 & 0.85 & maybe\\
\cline{2-11}
 & \multicolumn{2}{l||}{UAD} & 1.146 & 0.5 & 5.9 & 6.0 & 1.0 & 6 & 0.95 & \textbf{yes}\\
\specialrule{.1em}{.05em}{.05em}
\multirow{19}{.7cm}{GMV} &
 \multicolumn{2}{l||}{\textit{total} } & 0.331 & 0.6 & 3.3 & 7.2 & 2.2 & 12 & 0.77 & no\\
\cline{2-11}
 & \multirow{3}{.7cm}{CDS} & \textit{total} & 0.052 & 0.9 & 17.8 & 16.4 & 0.9 & 12 & 0.94 & \textbf{yes}\\
 &  & SNI & 0.031 & 0.6 & 17.5 & 17.2 & 1.0 & 12 & 0.94 & \textbf{yes}\\
 &  & MNI & 0.021 & 2.1 & 18.2 & 14.9 & 0.8 & 12 & 0.93 & \textbf{yes}\\
\cline{2-11}
 & \multirow{3}{.7cm}{FXD} & \textit{total} & 0.037 & 0.4 & 4.3 & 5.7 & 1.3 & 12 & 0.72 & no\\
 &  & F$\&$S & 0.013 & 0.5 & -9.9 & 6.4 & - & 0 & 0.66 & no\\
 &  & SWP & 0.017 & 0.5 & 2.3 & 5.2 & 2.2 & 12 & 0.67 & no\\
 &  & OPT & 0.006 & 0.5 & 7.0 & 7.4 & 1.1 & 12 & 0.88 & maybe\\
\cline{2-11}
 & \multirow{3}{.7cm}{IRD} & \textit{total} & 0.151 & 0.8 & -4.4 & 7.1 & - & 12 & 0.62 & no\\
 &  & FWD & 0.001 & 0.7 & 5.5 & 10.6 & 1.9 & 12 & 0.66 & no\\
 &  & SWP & 0.131 & 0.8 & -5.1 & 7.2 & - & 12 & 0.62 & no\\
 &  & OPT & 0.018 & 0.8 & 0.3 & 6.2 & 19.3 & 12 & 0.62 & no\\
\cline{2-11}
 & \multirow{3}{.7cm}{ELD} & \textit{total} & 0.019 & 0.6 & 7.3 & 8.0 & 1.1 & 0 & 0.95 & \textbf{yes}\\
 &  & F$\&$S & 0.005 & 0.5 & 8.2 & 7.8 & 0.9 & 6 & 0.93 & \textbf{yes}\\
 &  & OPT & 0.014 & 0.6 & 8.0 & 7.5 & 0.9 & 0 & 0.95 & \textbf{yes}\\
\cline{2-11}
 & \multirow{3}{.7cm}{CLD} & \textit{total}  & 0.036 & 0.9 & 10.0 & 15.8 & 1.6 & 0 & 0.86 & maybe\\
 &  & GLD & 0.001 & 0.5 & 5.2 & 4.4 & 0.8 & -6 & 0.68 & no\\
 &  & OTH & 0.035 & 1.2 & 10.6 & 16.7 & 1.6 & 0 & 0.87 & maybe\\
\cline{2-11}
 & \multicolumn{2}{l||}{UAD} & 0.037 & 0.5 & -2.6 & 7.2 & - & 6 & 0.81 & no\\
\cline{2-11}
 & \multicolumn{2}{l||}{GCE} & 0.063 & 0.4 & 4.5 & 4.1 & 0.9 & 12 & 0.85 & maybe\\
\specialrule{.1em}{.05em}{.05em}
\multicolumn{2}{|l|}{NOA-EXD} & FUT & 0.423 & 0.6 & 3.5 & 5.6 & 1.6 & -12 & 0.9 & \textbf{yes}\\
\multicolumn{2}{|l|}{\textcolor{white}{NOA-EXD}} &  OPT & 0.809 & 0.8 & 5.5 & 4.2 & 0.8 & -6 & 0.8 & no\\
\hline
\end{tabular}
\caption{\small{Fit results between the percolation edge density $\rho_p^{LD}$ of the LD-PIN and notional outstanding amounts (NOA) and gross-market values (GMV) of all major classes of OTC-traded financial derivatives and their first subcategories \cite{bis_der}, using the NLSMM (\ref{eq:mm}) (scaling factor $a_r$ and exponents $\gamma_{1,2}$ for a fit of $V_D(t)/V_r$). The fraction/multiple of world-GDP $f_{GDP}^{\mathrm{world}}$ of NOA/GMV as of the middle of 2008 is given for reference for each class/subcategory, where the mean between the 2008- and 2009-values has been taken for world-GDP. The quantity $m\equiv \gamma_2/\gamma_1$ indicates the time span over which changes in the LD-PIN and the derivative market are coupled (``market memory''), where a value greater than one means that past-year values of $\rho_p^{LD}$ contribute stronger in (\ref{eq:mm}) than present-year values. $\Delta t$ (in months) is the best-fit lead/lag time shift between $\rho_p^{LD}$ and NOA/GMV of a certain derivative class, where positive values indicate a lead of $\rho_p^{LD}$. We use the Pearson product correlation coefficient $p_r$ between the best-fit and market values of derivative products as a goodness-of-fit criterion, where we accept a fit if $p_r\geq 0.9\,$. We say that a certain product may be described by the NLSMM if $0.9>p_r\geq 0.85$ (conditional acceptance), and reject the fit if $p_r<0.85\,$. The major classes are credit default swaps (CDS), foreign exchange derivatives (FXD), interest rate derivatives (IRD), equity-linked derivatives (ELD) and commodity-linked derivatives (CLD). The first sub-categories are single- and multi- name instruments SNI and MNI for CDS, respectively, and forwards and swaps (F$\&$S), swaps (SWP), options (OPT) and forwards (FWD) for the other classes. CLD additionally include gold derivatives (GLD) and others commodities (OTH). Unallocated derivatives (UAD) are values which are not covered in \cite{bis_der}, but are included in \cite{bis_tri,isda_3}. Gross-credit exposure (GCE) measures the positive net-value of contracts, after mutual obligations have been set off (netting). Fit results for NOA of exchange-traded derivatives (EXD) are given for reference, where one sees that future contracts (FUT) can also be described by the NLSMM. The NLSMM \ref{eq:mm} is seen to be especially suitable for the description of the proliferation of CDS (NOA and GMV), which is the financial derivative product which has most-frequently been related to the GFC'08. Source:\ \cite{pin}.}}
}
\end{center}
\label{tab:der}
\end{table}
\end{document}